\documentclass[a4paper]{article}

\date{}
\usepackage[margin=1in]{geometry}
\usepackage{setspace}
\usepackage{times}
\usepackage{paralist}

\usepackage{graphicx} 
\usepackage{subfigure}

\usepackage{multirow}
\usepackage{booktabs}
\usepackage[hidelinks]{hyperref}

\usepackage{amsthm}
\newtheorem{example}{EXAMPLE}
\theoremstyle{definition}
\newtheorem{definition}{DEFINITION}

\usepackage{amssymb}
\usepackage{amsmath}

\begin{document}

\title{NLI4DB: A Systematic Review of Natural Language\\ Interfaces for Databases}

\author{Mengyi Liu and Jianqiu Xu\thanks{Corresponding author.} \\
	\normalsize \vspace{-0.5em} Nanjing University of Aeronautics and Astronautics, Nanjing, China \\
	\normalsize \{liumengyi,jianqiu\}@nuaa.edu.cn}

\maketitle

\begin{abstract}
As the demand for querying databases in all areas of life continues to grow, researchers have devoted significant attention to the \underline{n}atural \underline{l}anguage \underline{i}nterface for \underline{d}ata\underline{b}ases (NLIDB).
This paper presents a comprehensive survey of recently proposed NLIDBs.
We begin with a brief introduction to natural language processing techniques, executable database languages and the intermediate representation between natural language and executable language, and then provide an overview of the translation process from natural language to executable database language.
The translation process is divided into three stages: (i) \textit{natural language preprocessing}, (ii) \textit{natural language understanding}, and (iii) \textit{natural language translation}. 
Traditional and data-driven methods are utilized in the preprocessing stage.
Traditional approaches rely on predefined rules and grammars, and involve techniques such as regular expressions, dependency parsing and named entity recognition.
Data-driven approaches depend on large-scale data and machine learning models, using techniques including word embedding and pattern linking.
Natural language understanding methods are classified into three categories: (i) \textit{rule-based}, (ii) \textit{machine learning-based}, and (iii) \textit{hybrid}.
We then describe a general construction process for executable languages over relational and spatio-temporal databases.
Subsequently, common benchmarks and evaluation metrics for transforming natural language into executable language are presented, and methods for generating new benchmarks are explored.
Finally, we summarize the classification, development, and enhancement of NLIDB systems, and discuss deep language understanding and database interaction techniques related to NLIDB, including (i) \textit{using LLM for Text2SQL tasks}, (ii) \textit{generating natural language interpretations from SQL}, and (iii) \textit{transforming speech queries into SQL}.

\vspace{0.5em}
\noindent\textbf{Keywords: }Natural language interface for database, Semantic parsing, Structured language, Query processing

\end{abstract}

\section{Introduction}

In today's data-driven world, databases are the backbone of a number of applications, from social media platforms to financial systems. 
However, accessing and querying these vast repositories of information often requires specialized knowledge of query languages such as SQL, which can be a significant barrier for non-expert users, limiting their ability to harness the full potential of the data at their fingertips.
The advent of \underline{n}atural \underline{l}anguage \underline{i}nterface (NLI) has the potential to eliminate the interaction barrier between users and terminals \cite{Trummer21}.
The integration of \underline{n}atural \underline{l}anguage \underline{p}rocessing (NLP) and database technology represents an intriguing avenue for future research. 
There are systems that facilitate the transformation of natural language into structured language \cite{WangLKW24,ChangLWR24}, provide the natural language description for query execution plans \cite{WangBLJLC21,ChenLBJW22}, and transform SQL into natural language \cite{EleftherakisGK21,UstaKU24}.

\begin{figure}[!t]
	\centering
	\includegraphics[width=\linewidth]{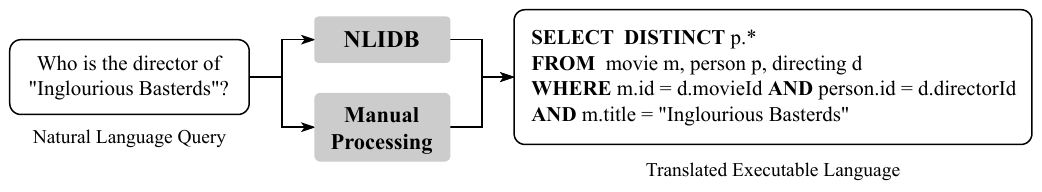}
	\caption{Example of translating a natural language into an executable language}
	\label{fig1}
\end{figure}

Imagine a world where anyone, regardless of technical proficiency, can effortlessly interact with complex databases using everyday language.
This vision is becoming a reality through the development of natural language interface for databases (NLIDB), which aims to transform a \underline{n}atural \underline{l}anguage \underline{q}uery (NLQ) into an executable language, as illustrated in Figure \ref{fig1}.
Users tend to favor an interactive interface that allows them to confirm the accuracy and precision of the generated structured language \cite{MokbelSXZAAAAAC22}.
The NLIDB enables users to avoid the necessity of possessing expertise in structured query languages and database schema, thereby significantly streamlining the efforts of users and enhancing the benefits of utilizing databases \cite{KimSHL20}.
The initial NLIDBs, including BASEBALL, LUNAR, LADDER, Chat-80, and ASK, were released in rapid succession \cite{AffolterSB19}. 
Subsequently, NLIDBs have emerged and are primarily utilized in relational databases (e.g., GENSQL \cite{FanRHWZL23} and CatSQL \cite{FuLWLTS23}), spatial domains (e.g., SpatialNLI \cite{SpatialNLI,WangLKW24} and NALSpatial \cite{LiuWX023}), RDF question and answer (e.g., Querix \cite{QUERIX} and TEQUILA \cite{JiaARSW18}), and XML databases (e.g., NaLIX \cite{LiYJ05,NALIX} and DaNaLIX \cite{LiCYSJ07}).

Despite years of research, the landscape of NLIDB is fraught with challenges \cite{BaikJ019,LiR17}. 
The inherent ambiguity and variability of natural language make NLIDB difficult to ensure accurate query interpretation.
Additionally, understanding the structure and semantics of different databases adds another layer of complexity.
Furthermore, achieving real-time performance while maintaining high accuracy in query translation remains an ongoing challenge.
While \underline{l}arge \underline{l}anguage \underline{m}odels (LLMs) offer new avenues for querying databases using natural language, the training and reasoning of such models necessitate a substantial amount of computational resources, which may prove challenging to implement in resource-limited scenarios \cite{Nan0ZRTZCR23}. 
Moreover, the decision-making process of LLMs is frequently opaque and lacks interpretability, making it difficult to ascertain whether the generated query results align with the user's intent \cite{SunZYGOCS23}.
These obstacles underscore the need for continued research and development to refine NLIDB.

\begin{figure}[!t]
	\centering
	\includegraphics[width=\linewidth]{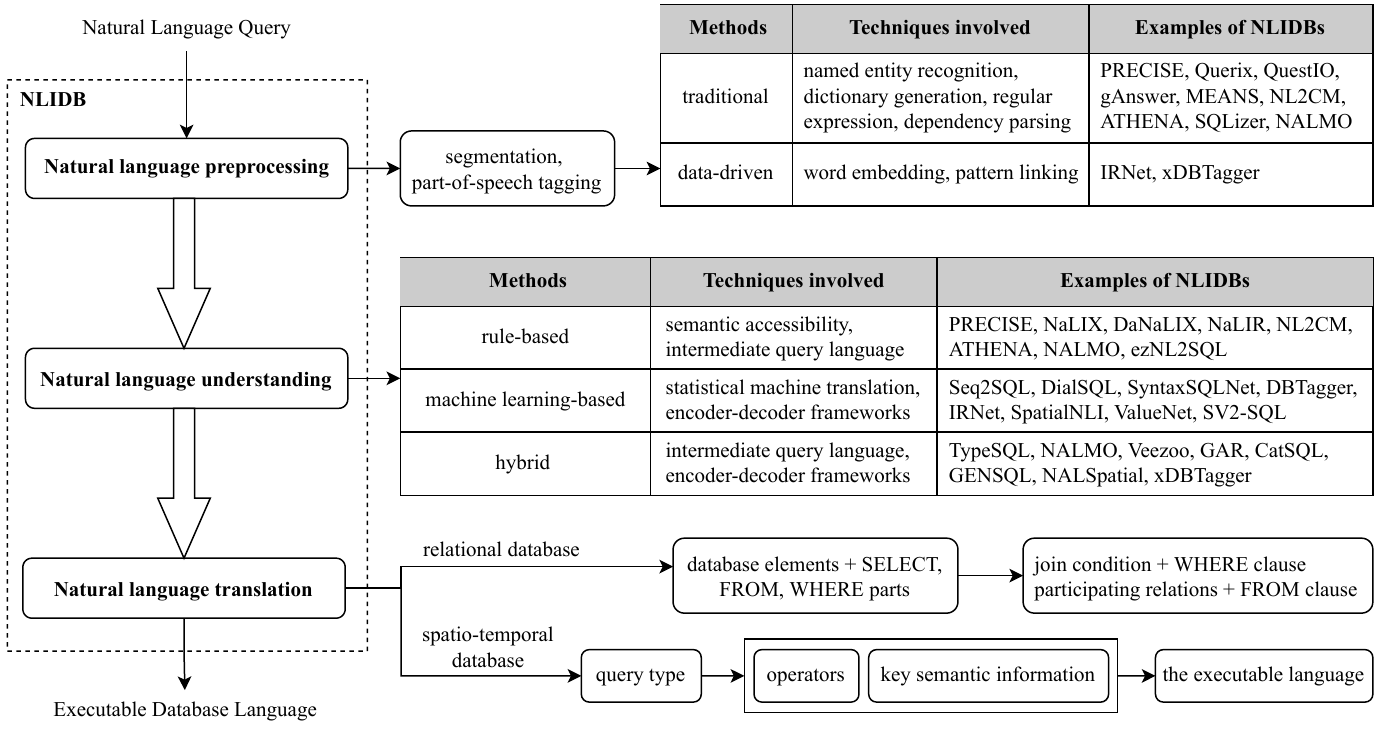}
	\caption{A summary of translation techniques}
	\label{fig2}
\end{figure}

In light of these observations, this systematic review explores the current state of NLIDB, examining the various approaches and technologies that have been proposed to connect natural language with database querying, named NLI4DB.
The aim of this survey is to offer a comprehensive overview that serves as both a valuable reference for researchers and a practical guide for practitioners aiming to implement effective NLIDB solutions.
NLI4DB presents a thorough examination of the NLIDB subject, categorizing the work into subtopics and providing in-depth analysis for each one.
The translation process from natural language to executable language is divided into three stages: (i) \textit{natural language preprocessing}, (ii) \textit{natural language understanding}, and (iii) \textit{natural language translation}. 
The three-stage division provides physical independence by separating the physical arrangement of data from the semantics of queries \cite{SahaFSMMO16}.
The techniques for the translation are shown in Figure \ref{fig2}.

(i) Natural language preprocessing generally involves the construction of dedicated data dictionaries for the domain using stemming extraction and synonym techniques.
Part-of-speech tagging and word segmentation are then performed on the input natural language.
Methods used in the preprocessing stage include traditional and data-driven.
Traditional approaches rely on predefined rules and grammars for domain-specific text processing, and involve techniques such as regular expressions, dependency parsing and \underline{n}amed \underline{e}ntity \underline{r}ecognition (NER).
Data-driven approaches depend on large-scale data and machine learning models for complex or variable text processing, using techniques such as word embedding and pattern linking.

(ii) Natural language understanding has rule-based, machine learning-based, and hybrid approaches.
Rule-based systems can only deal with knowledge bases of specific domains, whose semantic understanding processes either define the concept of semantic accessibility or translate the NLQ into an intermediate representation that can describe the semantics and relationships in an accessible manner.
Machine learning-based systems employ a variety of techniques to parse text, including unsupervised approaches, question-and-answer supervised learning, statistical machine translation techniques, encoder-decoder frameworks with recurrent neural networks, and combinations of deterministic algorithms and machine learning.
Hybrid approaches combine rules and machine learning techniques to maximize their benefits.

(iii) Natural language translation uses distinctive algorithms to map processed key semantic information to corresponding structured language components.
To build the SQL, the database elements matched by the NLQ are placed in the appropriate locations in the SELECT, FROM, and WHERE parts. 
In the event that a query involves multiple relations, it is necessary to include the join condition and the names of the participating relations in the WHERE and FROM clause, respectively.
In the process of building an executable language of a spatio-temporal database, the query type of the input NLQ is initially identified.
The operators required to build the executable language are then determined according to the query type. 
Finally, the key semantic information obtained from the natural language understanding stage is integrated to form an executable language.

The existing survey \cite{AffolterSB19} related to NLIDB focuses on the comparative analysis of the entire natural language interface system.
\textit{Affolter et al.} \cite{AffolterSB19} divide NLIs into four groups: (i) \textit{keyword-based systems}, (ii) \textit{pattern-based systems}, (iii) \textit{parsing-based systems}, and (iv) \textit{grammar-based systems}. 
For each group, they provide an overview of representative systems and describe the most illustrative one in detail.
In addition, they systematically compare 24 recently developed NLIDBs on the basis of the sample world designed in the paper. 
Each system is evaluated using 10 example questions to show the advantages and disadvantages.

Compared with \textit{Affolter et al.} \cite{AffolterSB19}, we divide the system translation process into three steps and focus on the comparative analysis of each step.
We investigate the recently developed NLIDB systems and divide the translation process into three stages: (i) \textit{natural language preprocessing}, (ii) \textit{natural language understanding}, and (iii) \textit{natural language translation}.
We classify natural language preprocessing techniques into traditional and data-driven.
Natural language understanding methods are then analyzed in three categories: (i) \textit{rule-based}, (ii) \textit{machine learning-based}, and (iii) \textit{hybrid}.
Next, we provide a comprehensive outline of the construction process of executable languages for relational and spatio-temporal databases. 
Finally, we present commonly used benchmarks and evaluation metrics, and describe the classification, development, and enhancement of NLIDBs.

The rest of the paper is structured as follows. 
Section \ref{sec2} furnishes the background concerning NLIDB, including natural language processing techniques, executable database languages and intermediate representation languages. 
Section \ref{sec3} describes the generation of executable database languages in terms of three stages: (i) \textit{natural language preprocessing}, (ii) \textit{natural language understanding} and (iii) \textit{natural language translation}. 
Section \ref{sec4} summarizes 11 popular benchmarks for transforming NLQ into SQL and 3 evaluation metrics, including response time, translatability, and translation precision, and explores the methods for generating new benchmarks. 
Section \ref{sec5} analyzes the classification, development and enhancement of NLIDBs.
Section \ref{sec6} discuss deep language understanding and database interaction techniques related to NLIDB, including (i) \textit{using LLM for Text2SQL tasks}, (ii) \textit{generating natural language interpretations from SQL}, and (iii) \textit{transforming speech queries into SQL}.
Section \ref{sec7} explores the open problems of NLIDB and concludes the survey.
Table \ref{tab1} summarizes the frequently used notations.

\begin{table}[!t]
	\centering
	\caption{Frequently used notations}
	\label{tab1}
	\begin{tabular}{lc}
		\toprule
		\textbf{Name} & \textbf{Abbreviation} \\
		\midrule
		Natural language interface for database & NLIDB \\
		Natural language interface & NLI \\
		Natural language query & NLQ \\
		Natural language processing & NLP \\ 
		Named entity recognition & NER \\
		Large language model & LLM \\
		First-order logic & FOL \\
		Automatic speech recognition & ASR \\
		sequence-to-sequence & seq2seq \\
		\bottomrule
	\end{tabular}
\end{table}

\section{Background: NLP techniques and query languages}\label{sec2}

We introduce the background related to NLIDB, including natural language processing techniques, executable database languages, and intermediate representation languages.

\subsection{Natural language processing techniques}

NLP is an interdisciplinary discipline that integrates several fields such as linguistics, computer science, and mathematics, and aims to make computers capable of understanding, processing and generating natural language text or speech.
Through segmentation, lexical annotation, and syntactic analysis, NLP provides structured processing of text to achieve semantic understanding and information extraction. 
The application areas of NLP cover machine translation, sentiment analysis, information retrieval, and dialogue systems, providing people with an intelligent and convenient way of language interaction.

\textbf{A Brief History.}
The earliest research on natural language processing is machine translation.
In 1950, Alan Turing proposed the ultimate test for determining the arrival of truly ``\textit{intelligent}" machines, which is generally regarded as the inception of the idea of NLP \cite{abs-1209-6238}.
From the 1950s to the 1970s, the rule-based method was used to process natural language, which was based on grammatical rules and formal logic.
In the 1970s, the statistic-based method gradually supplanted the rule-based method.
At this juncture, NLP built on mathematical models and statistic made a substantial breakthrough and was applied to practical applications.
From 2008 to the present, researchers have introduced deep learning to NLP in response to the achievements in image recognition and speech recognition.

The NLP techniques commonly used in NLIDBs are as follows.

\begin{figure}[!t]
	\centering
	\subfigure[Part of speech tagging]{
		\label{fig3-1}
		\includegraphics[width=0.5\textwidth]{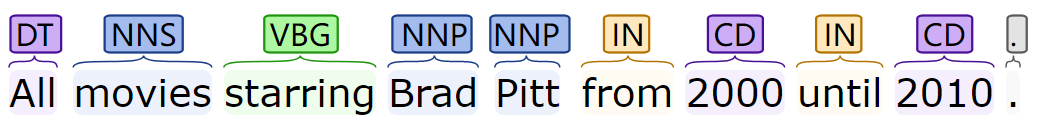}}
	\\
	\subfigure[Lemmatization]{
		\label{fig3-2}
		\includegraphics[width=0.5\textwidth]{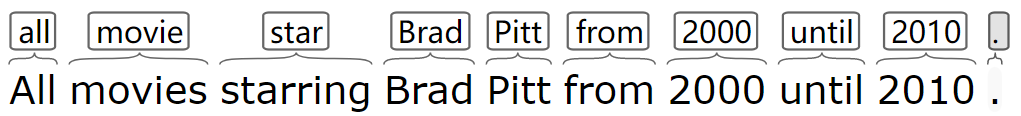}}
	\\
	\subfigure[Named entity recognition]{
		\label{fig3-3}
		\includegraphics[width=0.5\textwidth]{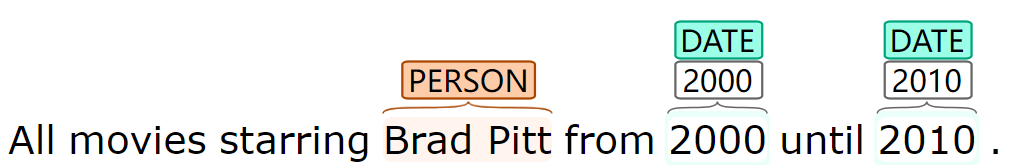}}
	\\
	\subfigure[Dependency parsing]{
		\label{fig3-4}
		\includegraphics[width=0.8\textwidth]{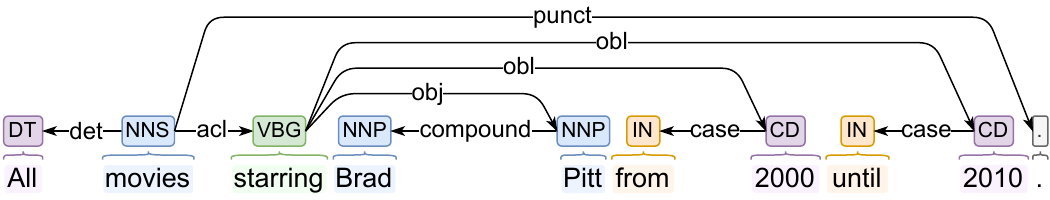}}
	\caption{Processing natural language using Stanford CoreNLP}
	\label{fig3}
\end{figure}

(i) \textbf{Part of speech tagging} refers to assigning the correct part of speech to each word in the segmented text, determining whether each word is a noun, verb, or adjective. 
In NLIDB, part of speech tagging facilitates the identification of the grammatical roles of individual words in natural language queries, leading to an accurate comprehension of users' intent.
Taking the natural language query ``\textit{All movies starring Brad Pitt from 2000 until 2010.}" as an example, the result of part of speech tagging using Stanford CoreNLP is shown in Figure \ref{fig3-1}.
In the figure, DT = determiner; NNS = plural noun; VBG = the gerund or present participle of a verb; NNP = singular proper noun; IN = preposition or subordinating conjunction; CD = cardinal number.

(ii) \textbf{Lemmatization} is the process of reducing the different forms of a word to the original form. 
In NLIDB, lemmatization is beneficial in unifying words of various tenses and morphs in natural language queries into base forms in order to match the content in the database.
Taking the natural language query ``\textit{All movies starring Brad Pitt from 2000 until 2010.}" as an example, the result of lemmatization using Stanford CoreNLP is shown in Figure \ref{fig3-2}.

(iii) \textbf{Named entity recognition} is the procedure of identifying entities with specific meanings in natural language text \cite{SchmittKRPT19}. 
Generally, the recognized entities can be categorized into three primary groups (entity, temporal, and numeric) and seven subgroups (PERSON, ORGANIZATION, LOCATION, TIME, DATE, MONEY, and PERCENT). 
NER in NLIDB enables the identification of entities involved in a natural language query to locate the topic and scope of the query.
Taking the natural language query ``\textit{All movies starring Brad Pitt from 2000 until 2010.}" as an example, the result of NER using Stanford CoreNLP is shown in Figure \ref{fig3-3}.

(iv) \textbf{Dependency parsing} involves analyzing the dependencies between words in natural language sentences.
A binary asymmetric relationship between words is called dependency, which is described as an arrow from the head (the subject to be modified) to the dependent (the modifier). 
Dependency parsing in NLIDB facilitates the understanding of grammatical relationships between words in NLQs, so that the structure and meaning of the query can be accurately understood.
Taking the natural language query ``\textit{All movies starring Brad Pitt from 2000 until 2010.}" as an example, the result of dependency parsing using Stanford CoreNLP is illustrated in Figure \ref{fig3-4}.
In the figure, punct = punctuation; obl = oblique nominal; obj = object; det = determiner; acl = clausal modifier of noun; case = case marking.

With the vigorous development of NLP technology, a number of NLP tools are appearing \cite{SchmittKRPT19}. These tools can perform basic tasks, including dependency parsing, named entity recognition, lemmatization, and part of speech tagging, each of which has distinct advantages and disadvantages. 
The following is a list of the established open source natural language processing tools.

(i) \textbf{NLTK} is a natural language processing toolkit using Python as the programming language. 
NLTK has complete functions and realizes many of the functional components in natural language processing, such as named entity recognition, sentence structure analysis, part-of-speech tagging, and text classification \cite{Bird06}. 
Born for the academic field, NLTK is suitable for study and research. 
The disadvantage is that NLTK has a slower processing speed than other tools.

(ii) \textbf{spaCy}, a commercial open source software, is an industrial-grade natural language processing software programmed in Python and Cython languages \cite{FantechiGLS21}. 
spaCy, which follows NLTK, includes pre-trained statistical models and word vectors. 
spaCy can break down text into semantic units like articles, words and punctuation, and support named entity recognition.
spaCy is characterized by fast and accurate syntax analysis, and comprehensive functions ranging from simple part-of-speech tagging to advanced deep learning.

(iii) \textbf{Stanford CoreNLP} is a tool set developed by Stanford University using the Java programming language.
Stanford CoreNLP supports a variety of natural languages and has rich interfaces for programming languages that can be used without Java \cite{ManningSBFBM14}. 
Stanford CoreNLP is an efficient tool created by high-level research institutions and is widely used in scientific research and experiments, but may incur additional costs in production systems. Stanford CoreNLP may not be the best choice for industry.

(iv) \textbf{TextBlob} is an extension to NLTK, which provides an easier way to use the functionality of NLTK \cite{HazarikaKDB20}. TextBlob supports sentiment analysis, tokenization, part-of-speech tagging, and text classification. 
One of the advantages is that TextBlob can be used in production environments where performance requirements are not too high.
TextBlob can be applied in a wide range of scenarios, especially for small projects.

\subsection{Executable database languages}

The output of NLIDB is an executable database language, and we present executable languages over relational data, RDF data, and spatial data.

\subsubsection{Query language for relational data}

The standard executable query language for relational data is SQL. 
Such a language is a general-purpose, extremely powerful relational database language whose functions are not limited to querying, but also include creating database schema, inserting and modifying data, and defining and controlling database security integrity \cite{Date84a}. 
Following the establishment of SQL as an international standard language, numerous database manufacturers have released SQL-compatible software, including both database management systems and interfaces.
Consequently, SQL serves as the universal data access language and standard interface for most databases, fostering a shared foundation for interoperability among different database systems.
SQL has become the mainstream language in the database field which is of great significance.

SQL provides the SELECT statement for querying data, which has flexible usage and rich functionality. 
The SELECT statement can perform simple single-table queries as well as complex join queries and nested queries, whose general format is:

%\vspace{0.5em}
\begin{flushleft}
	\textit{\textbf{SELECT} [ALL$|$DISTINCT] $<$target column expression$>$ [alias] [,$<$target column expression$>$ [alias]]}
	
	\textit{\textbf{FROM} $<$table name or view name$>$ [,$<$table name or view name$>$] $|$ (SELECT statement) [AS] $<$alias$>$}
	
	\textit{[\textbf{WHERE} $<$conditional expression$>$]}
	
	\textit{[\textbf{GROUP BY} $<$column name 1$>$ [\textbf{HAVING} $<$conditional expression$>$]]}
	
	\textit{[\textbf{ORDER BY} $<$column name 2$>$ [ASC$|$DESC]];}
\end{flushleft}
%\vspace{0.5em}

\noindent The purpose of the SELECT statement is to find the tuples that satisfy the conditions specified in the FROM clause, which may be a basic table, view, or derived table ,according to the conditional expression in the WHERE clause. 
The attribute value in the tuple is then selected on the basis of the target column expression in the SELECT clause to form the result table. 
When a GROUP BY clause is present, the output is organized by the value of $<$\textit{column name 1}$>$, where tuples sharing identical attribute column values are grouped together.
Aggregation functions are usually applied to each group. 
When the GROUP BY clause is accompanied by a HAVING clause, the output will only include groups that satisfy the specified conditions.
If an ORDER BY clause is present, the result table is sorted in ascending or descending order according to the values of $<$\textit{column name 2}$>$.

\subsubsection{Query language for RDF data}

The complete designation of RDF is Resource Description Framework, which is a data model designed to represent information about resources on the Internet.
The data model typically describes a fact composed of three parts known as a triple, including (i) \textit{a subject}, (ii) \textit{a predicate}, and (iii) \textit{an object}.
An RDF graph contains multiple triples. RDF documents are written in XML to offer a standardized method for describing information. 
RDF is intended for computer applications to read and understand, rather than for visual presentation to web users.

SPARQL is a specialized query language and data retrieval protocol designed for RDF, which stands for SPARQL Protocol and RDF Query Language \cite{ChenLO21}. 
SPARQL is a query language over RDF graphs, where the database is represented as a collection of ``\textit{subject-predicate-object}" triples.
Although RDF data is inferential, SPARQL does not have an inference query function. 
SPARQL is tailored for managing data stored in RDF format, enabling both retrieval and manipulation.
SPARQL is composed of the following components.
\begin{compactitem}
	\item The PREFIX clause is employed to declare a prefix with the objective of simplifying the use of URIs. The declaration of the prefix is optional.
	\item The SELECT clause serves the purpose of specifying the variables returned by a query.
	\item The WHERE clause is utilized to match data in RDF graphs. The clause contains one or more triple patterns that are employed to indicate the conditions of a query.
	\item The FILTER clause is designed to conditionally filter the results of a query. The clause can include boolean expressions to limit the set of results matched by the WHERE clause.
\end{compactitem}

The fundamental query types of SPARQL are as follows \cite{Parsia06}.

\textbf{SELECT query} is the most frequently used type of query, whose function is to select variables and return a result set.
A table is typically generated as the outcome of a SELECT query, which includes the variables that meet the query's criteria along with their corresponding values.

\textbf{CONSTRUCT query} is used to generate a new RDF graph by utilizing the query pattern.
In contrast to tabular results, the CONSTRUCT query produces an RDF graph that is constructed from the matching data of the query pattern.

\textbf{ASK query} is designed to ascertain the existence of RDF data that satisfies the query pattern. The ASK query provides a response in the form of a boolean value (true or false) to indicate the presence or absence of a match.

\textbf{DESCRIBE query} is employed to obtain the detailed description of resources. The description is determined by the query engine and typically consists of triples that are directly related to the resource.

Each query type employs a WHERE clause to limit the scope of the query. Nevertheless, in the context of DESCRIBE queries, the inclusion of a WHERE clause is not mandatory. To illustrate, the subsequent query retrieves people from the data set who are above the age of 24:

%\vspace{0.5em}
\begin{flushleft}
	\textit{PREFIX info: $<$http://somewhere/peopleInfo\#$>$}
	
	\textit{SELECT ?resource}
	
	\textit{WHERE}
	
	\textit{\{}
	
	\textit{\hspace{1em}?resource info:age ?age .}
	
	\textit{\hspace{1em}FILTER (?age $>=$ 24)}
	
	\textit{\}}
\end{flushleft}
%\vspace{0.5em}

\noindent In this query, the ``\textit{?}" symbol represents a variable, followed by the variable name. The middle of the ``$ <> $" symbol is the URI that describes the resource address. 
The ``\textit{info:age}" in the above query is a URI shorthand and stands for ``$ < $\textit{http://somewhere/peopleInfo$\#$age}$ > $". 
The FILTER keyword is employed to impose limitations on the outcomes that are retrieved.
In addition, RDF is semi-structured data, and different entities in RDF may have distinct properties. 
SPARQL is capable of querying information that exists in RDF.
However, when querying information that does not exist, SPARQL does not show a failure and does not return any results. 
The OPTIONAL keyword can then be used to signify that the query is optional, indicating that the query will return a result if the entity has the attribute, and a null value otherwise.
The FILTER keyword can also be used in conjunction with the OPTIONAL keyword.

\subsubsection{Query language for spatial data}

The increasing reliance on geographic information systems in many aspects of people's production and life has led to a significant increase in the demand for spatial data query in all walks of life. 
The popularity of spatial applications has brought great attention to spatial databases \cite{Guting94}. 
In databases, fundamental data types utilized for the representation and manipulation of spatial objects include point, line, and region.
The common operators to query spatial data are shown in Table \ref{tab2}.

\begin{table}[!t]
	\caption{Operators to query spatial data}
	\label{tab2}
	\centering
	\begin{tabular}{lp{0.46\textwidth}p{0.345\textwidth}}
		\toprule
		\textbf{Operator} & \textbf{Signature} & \textbf{Meaning} \\
		\midrule
		%\hline
		distance & point $|$ line $|$ region $ \times $ point $|$ line $|$ region $\rightarrow$ real & Compute the distance between two spatial objects. \\
		%\hline
		direction & point $\times$ point $\rightarrow$ real & Compute the direction between two points. \\
		%\hline
		size & line $\rightarrow$ real & Return the length of a line. \\
		%\hline
		area & region $\rightarrow$ real & Return the area of a region. \\
		%\hline
		intersects & line $|$ region $ \times $ line $|$ region $ \rightarrow$ bool & TRUE, if both arguments intersect. \\
		%\hline
		intersection & point $|$ line $|$ region $ \times $ point $|$ line $|$ region $\rightarrow$ T, where T is point if point is one of the arguments, otherwise T is the argument having the smaller dimension  & Intersection of two spatial objects. \\
		%\hline
		distancescan & rtree $\times$ relation $\times$ object $\times$ int $\rightarrow$ stream & Compute the integer \textit{k} nearest neighbors for a query object. \\
		%\hline
		\bottomrule
	\end{tabular}
\end{table}

Mature systems for storing and managing spatial data include Esri's ArcGIS, PostGIS, Google Earth Engine, GRASS GIS, and SECONDO \cite{GutingBD10}.
As an illustration, SECONDO is a freely available platform created for the purpose of organizing and examining spatial and temporal data.
The basic commands of SECONDO are as follows.

\textbf{query $<$value expression$>$.}
The command evaluates the given value expression and subsequently displays the result to the user.

\textbf{let $<$identifier$>$ = $<$value expression$>$.}
The command initially evaluates the provided value expression in a manner analogous to the preceding command.
In contrast to the previous command, the results of the evaluation are not immediately displayed but rather stored in an object named \textit{identifier}. 
If the object already exists in the database, the command will result in an error.

\textbf{delete $<$identifier$>$.}
The command removes the object named \textit{identifier} from the current database, and is typically utilized in conjunction with the second command.

\begin{figure}[!t]
	\centering
	\includegraphics[width=\textwidth]{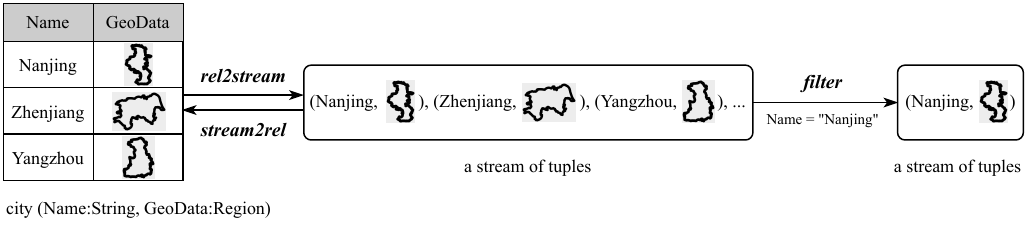}
	\caption{Functions of the operators rel2stream, stream2rel and filter in SECONDO}
	\label{fig4-2}
\end{figure}

%SECONDO's executable query language is considered to be a more rudimentary form of language in comparison to SQL. A comprehensive comprehension of the intricate relationship between data flow and operators is imperative when writing.
When an expert or system developer writes the executable query language for SECONDO, one needs to comprehensively understand the intricate relationship between data flow and operators.
The \textit{rel2stream} operator transforms a relation into a stream of tuples, as shown in Figure \ref{fig4-2}.
The \textit{stream2rel} operator, in contrast, converts a stream of tuples into a relation.
Among the fundamental operators of SECONDO, the \textit{filter} operator is the most frequently utilized.
Similar to the SELECT keyword in SQL, the function of the \textit{filter} operator is to extract information from data that satisfies specific conditions.
The SELECT keyword operates on a two-dimensional table structure to query, filter, and project data by specifying columns and conditions.
The \textit{filter} operator works on a stream of tuples, followed by a filter condition. 
The tuples that match the condition are then collected and outputted as a stream.
For example, the following executable language will output all information about Nanjing in the relation \textit{city} in SECONDO. 

%\vspace{0.5em}
\begin{flushleft}
	\textit{query city rel2stream filter [.Name = ``Nanjing"] stream2rel;}
\end{flushleft}
%\vspace{0.5em}

\noindent 
During the execution of the query, the \textit{rel2stream} operator first transforms the relation \textit{city} into a stream of tuples, then the \textit{filter} operator extracts the tuple named ``\textit{Nanjing}" from the stream, and finally the \textit{stream2rel} operator converts the tuple into a relation from the stream.

\subsection{Intermediate representation languages}

The intermediate representation language in NLIDB is designed to accommodate the semantic discrepancies and diversity between natural language and executable database language, thus improving the translation accuracy, flexibility and maintainability of the system \cite{AndroutsopoulosRT95}.
The intermediate representation serves as a translator between natural language and executable language, mapping complex natural language structures to a unified semantic representation for the purpose of efficient subsequent query processing and execution.
By decoupling NLQ from the underlying database query language, the intermediate representation language makes the NLIDB system flexible, portable, and adaptable to various database types and query requirements.
The design of the intermediate representation considers several factors, such as:

(i) The intermediate representation should convey the query request that the user wishes to submit to the database, rather than the full meaning of the user's input.

(ii) To facilitate subsequent translation into the executable language of the database, the intermediate representation should be unambiguous. 

(iii) To make re-development easier, the intermediate representation should be reusable.

Popular intermediate representations are parse trees \cite{LiJ14SIGMOD}, first-order logic \cite{SongSSBZBMDDMH15}, OQL \cite{SahaFSMMO16}, query sketch \cite{Yaghmazadeh0DD17}, SemQL \cite{GuoZGXLLZ19}, and NatSQL \cite{GanCXPWDZ21}.

\textbf{Parse tree.} 
The syntactic structure of a query in natural language is closely tied to the design of a parse tree.
The tree structure is typically applied to represent the hierarchical and structural relationships of the query.
Each node in a parse tree indicates a grammatical unit (e.g., phrase, word group, and vocabulary), while edges indicate grammatical relations (e.g., modification and conjunction) between these grammatical units.
The nodes and edges on the parse tree can be labeled with semantic information to identify the semantic roles and constraints present in the query, providing important information for subsequent query processing.

\textbf{First-order logic.} When transforming an NLQ into \underline{f}irst-\underline{o}rder \underline{l}ogic (FOL), words and phrases in the natural language are first mapped to predicates, constants, variables, and logical connectives in FOL to represent entities, attributes, and relations in the query.
Subsequently, on the basis of the syntactic structure of the NLQ, the syntax tree or syntax graph of the FOL representation is constructed to capture the semantic relations and logical structures in the query.
Finally, the topics, conditions, and operations in the query are identified and converted into logical expressions in FOL to denote the constraints and operational requirements of the query.

When converting the natural language query ``\textit{Find the names and salaries of all employees older than 30.}" into a first-order logic representation, predicates and constants are defined as follows.

$ Employee(x) $: \textit{x} is an employee

$ Name(x, n) $: the name of employee \textit{x} is \textit{n}

$ Age(x, a) $: the age of employee \textit{x} is \textit{a}

$ Salary(x, s) $: the salary of employee \textit{x} is \textit{s}

The query condition is expressed as $ \forall x (Employee(x) \wedge Age(x,a) \wedge a > 30) $. The query result is expressed as $ \exists n, s (Name(x, n) \wedge  Salary(x, s)) $. 
The complete first-order logic representation is obtained by combining the condition and result of the query.
\begin{equation}
	\forall x (Employee(x) \wedge Age(x,a) \wedge a > 30) \wedge \exists n, s (Name(x, n) \wedge  Salary(x, s))
	\nonumber
\end{equation}

\textbf{OQL} is built on an ontology knowledge graph, where words and phrases in natural language queries are associated with concepts, attributes and relations within the ontology knowledge graph.
The semantic information of natural language queries is captured through semantic representations and query patterns to effectively interact with the database.
OQL grammars permit the expression of complex aggregation, union and nested queries. 
OQL queries operate upon individual concepts, with each concept being assigned an alias as specified in the FROM clause of the query.

\textbf{Query sketch} is a form of SQL with natural language hints.
Taking the NLQ ``\textit{Find the number of papers in OOPSLA 2010.}" as an example, the query sketch is as follows.
%\vspace{0.5em}
\begin{flushleft}
	\textit{SELECT count(?[papers]) FROM ??[papers] WHERE ? = ``OOPSLA 2010";}
\end{flushleft}
%\vspace{0.5em}

\noindent In the query sketch, the symbols ``\textit{??}" and ``\textit{?}" represent an unspecified table and an unspecified column, respectively.
Hints for the corresponding gaps are indicated by words enclosed in square brackets.
As an illustration, the first hint in the sketch suggests that the symbol ``\textit{?}" has a similar semantic meaning to the term \textit{papers}.

\textbf{SemQL} is designed as a tree structure that not only constrains the search space during synthesis, but also maintains the same structural characteristics as SQL. 
In SemQL queries, the GROUP BY, HAVING, and FROM clauses in SQL are removed, and the conditions from the WHERE and HAVING clauses are consistently represented in the Filter sub-tree.
Furthermore, in the later inference phase, domain knowledge is utilized to deterministically infer implementation details from SemQL queries.
For instance, the columns included the GROUP BY clause of SQL are typically present in the SELECT clause.

\textbf{NatSQL} retains the core functionality of SQL while streamlining the structure of SQL to align more closely with the syntax of natural language.
NatSQL keeps only the SELECT, WHERE, and FROM clauses, omitting the JOIN ON, HAVING, and GROUP BY clauses.
Additionally, NatSQL does not require nested sub-queries or aggregation operators, and employs a single SELECT clause.
In the case of the natural language query ``\textit{Which film has more than 5 actors and less than 3 in the inventory?}", the SQL and NatSQL are as follows.
%\vspace{0.5em}
\begin{flushleft}
	\textit{\textbf{SQL:} SELECT T1.title FROM film AS T1 JOIN film\_actor AS T2 ON T1.film\_id = T2.film\_id GROUP BY T1.film\_id HAVING count(*) $>$ 5 INTERSECT SELECT T1.title FROM film AS T1 JOIN inventory AS T2 ON T1.film\_id = T2.film\_id GROUP BY T1.film\_id HAVING count(*) $<$ 3;}
	
	\textit{\textbf{NatSQL:} SELECT film.title WHERE count(film\_actor.*) $>$ 5 and count(inventory.*) $<$ 3;}
\end{flushleft}
%\vspace{0.5em}

\section{Generation of executable database languages}\label{sec3}

The generation of executable database languages can be divided into three stages: (i) \textit{natural language preprocessing}, (ii) \textit{natural language understanding}, and (iii) \textit{natural language translation}.
In stage (i), the system performs a preliminary analysis of the raw natural language query in order to prepare for the subsequent stage of natural language understanding.
In stage (ii), the system performs semantic parsing and understanding of the preprocessed natural language query to extract the semantic details and intent of the query.
In stage (iii), the system converts the comprehended natural language into a language that can be executed within the database.

\subsection{Natural language preprocessing}

Prior to the semantic understanding and translation of natural language queries, preprocessing is performed using traditional and data-driven methods.
In order to preprocess natural language queries, the recently developed NLIDBs utilize techniques as illustrated in Table \ref{tab3}.

\begin{table}[!t]
	\centering
	\caption{Natural language preprocessing for NLIDBs}
	\label{tab3}
	\begin{tabular}{l|l|l|c|c|c|c|c|c|c|c}
		\hline
		\textbf{NLIDB} & \textbf{Year} & \textbf{Underlying datatype} & \rotatebox{90}{\textbf{Segmentation}} & \rotatebox{90}{\textbf{Part of speech}} & \rotatebox{90}{\textbf{NER}} & \rotatebox{90}{\textbf{Dictionary generation}} & \rotatebox{90}{\textbf{Regular expression}} & \rotatebox{90}{\textbf{Dependency parsing}} & \rotatebox{90}{\textbf{Word Embedding}} & \rotatebox{90}{\textbf{Pattern Linking}} \\
		\hline
		PRECISE \cite{PopescuEK03} & 2003 & relational data & \checkmark & \checkmark & \checkmark & \checkmark &  & & & \\
		\hline
		Querix \cite{QUERIX} & 2006 & ontology & \checkmark & \checkmark &  & \checkmark & & & & \\
		\hline
		QuestIO \cite{DamljanovicTB08} & 2008 & ontology & \checkmark & \checkmark &  & \checkmark & & & & \\
		\hline
		gAnswer \cite{HuangZ13} & 2013 & RDF data & & & & \checkmark & & & & \\
		\hline
		MEANS \cite{AbachaZ15} & 2015 & RDF data & \checkmark & & \checkmark & & & & & \\
		\hline
		NL2CM \cite{AmsterdamerKM15SIGMOD,AmsterdamerKM15VLDB} & 2015 & RDF data & \checkmark & \checkmark & & & & \checkmark  & & \\
		\hline
		NL2TRANQUYL \cite{BoothECW15} & 2015 & relational data & & & & & & \checkmark & & \\
		\hline
		ATHENA \cite{SahaFSMMO16} & 2016 & relational data & \checkmark & \checkmark & \checkmark & & & \checkmark & & \\
		\hline
		SQLizer \cite{Yaghmazadeh0DD17} & 2017 & relational data & \checkmark & \checkmark & \checkmark & & & & & \\
		\hline
		TEQUILA \cite{JiaARSW18} & 2018 & RDF data & \checkmark & \checkmark & \checkmark & & & & & \\
		\hline
		MyNLIDB \cite{DasB29} & 2019 & relational data & \checkmark & \checkmark & & & & \checkmark & & \\
		\hline
		IRNet \cite{GuoZGXLLZ19} & 2019 & relational data & & & & & & & \checkmark & \checkmark \\
		\hline
		NLMO \cite{WangXW20} & 2020 & moving objects & \checkmark & \checkmark & \checkmark & \checkmark & \checkmark & & & \\
		\hline
		NALMO \cite{WangX021,WangLXL23} & 2021 & moving objects & \checkmark & \checkmark & \checkmark & \checkmark & \checkmark & & & \\
		\hline
		NALSD \cite{LiuWX23} & 2023 & spatial data & \checkmark & \checkmark & \checkmark & \checkmark & & & & \\
		\hline
		NALSpatial \cite{LiuWX023} & 2023 & spatial data & \checkmark & \checkmark & \checkmark & \checkmark & & & & \\
		\hline
		xDBTagger \cite{UstaKU24} & 2024 & relation data & \checkmark & \checkmark &  & & &  & \checkmark & \checkmark \\
		\hline
	\end{tabular}
\end{table}

The preprocessing process of many NLIDBs commences with the construction of a dedicated data dictionary for the domain. The extraction process of domain knowledge exerts a profound influence on the portability of the system. 
In addition, the semantic parsing component needs to accurately comprehend NLQ with the assistance of the dictionary, and the extraction process of domain knowledge will impact the availability of the NLIDB. 
The primary goal of the extraction technique is to minimize the burden on system users while enhancing the capacity to automatically generate a dictionary. 
The extraction process is primarily reliant on stemming and synonym techniques. 
The system then needs to perform word segmentation and part-of-speech tagging on the input natural language. 
This process necessitates the utilization of natural language processing tools. 
When choosing the tool, the high accuracy of the segmentation and part-of-speech tagging results should be considered first, followed by the speed of processing.
Furthermore, the query must be oriented to database information, and the relevant statements used in the query request are closely related to the database to be used. 
Therefore, part-of-speech tagging is often employed in conjunction with named entity recognition and data dictionary.

%\begin{figure}[!t]
%	\centering
%	\includegraphics[width=0.9\textwidth]{fig2-2-1.pdf}
%	\caption{The technologies involved in preprocessing}
%	\label{fig2-2-1}
%\end{figure}

Traditional preprocessing methods rely on predefined rules and grammars, involving techniques including NER, regular expressions, and dependency parsing.
NLMO performs segmentation and entity recognition using a natural language processing toolkit spaCy, and sets regular expressions for temporal information extraction.
ATHENA utilizes the TIMEX annotator to detect all temporal intervals mentioned in the text, and the Stanford Numeric Expressions annotator to pinpoint all tokens containing numerical values. 
ATHENA employs the Stanford Dependency Parser to identify the dependency relationship in the context of the GROUP BY clause.
PRECISE utilizes the Charniak parser for the precise parsing of questions and the extraction of token relationships from the resulting parse tree.
NL2CM employs dependency parsing and part-of-speech tagging techniques.
NL2TRANQUYL analyzes the input natural language using the Stanford Parser, resulting in constituency and dependency parses. 

Data-driven preprocessing methods depend on large-scale data and machine learning models, and the techniques used include word embedding and pattern linking.
Word2Vec and GloVe are word embedding models that are able to represent words as points in a sequential vector space, thereby capturing the semantic relationships between words.
These vectors can be employed for calculating semantic similarity and extracting features.
xDBTagger utilizes a pre-trained word embedding model to convert tokens into a 300-dimensional vector representation.
IRNet performs schema linking by connecting the natural language with the database schema, aiming to identify the specific columns and tables referenced in the natural language.
The columns are then assigned different types according to the manner mentioned in the question.

\subsection{Natural language understanding}

\begin{figure}[!t]
	\centering
	\subfigure[Rule-based methods]{
		\label{fig5-1}
		\includegraphics[width=\textwidth]{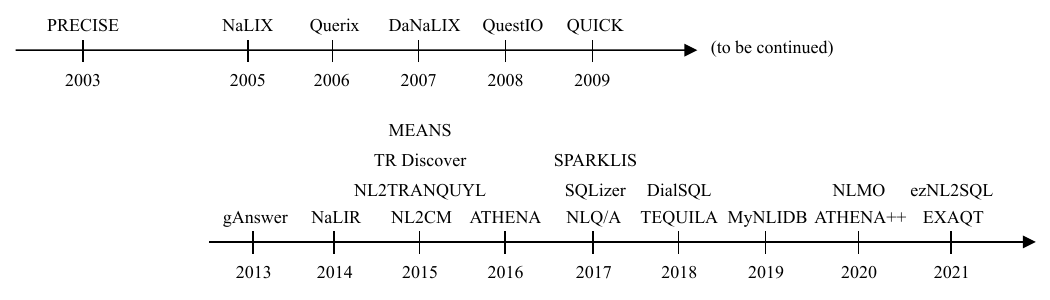}}
	\\
	\subfigure[Machine learning-based methods]{
		\label{fig5-2}
		\includegraphics[width=\textwidth]{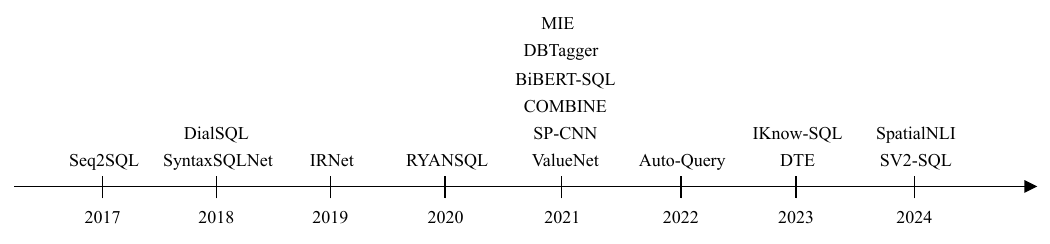}}
	\\
	\subfigure[Hybrid methods based on rule and machine learning]{
		\label{fig5-3}
		\includegraphics[width=\textwidth]{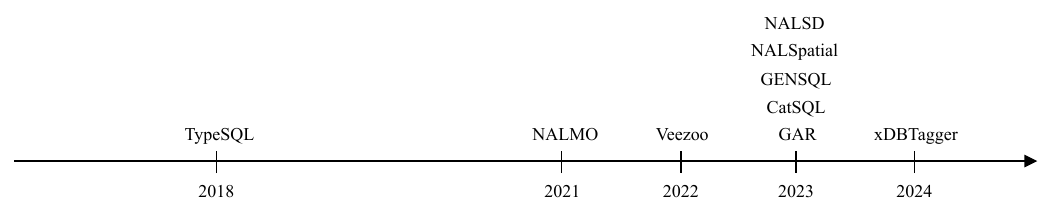}}
	\caption{Timelines of the research progress of techniques for understanding natural language}
	\label{fig5}
\end{figure}

Three principal technical approaches to understand natural language are (i) \textit{rule-based}, (ii) \textit{machine learning-based}, and (iii) \textit{hybrid}.
Based on the techniques, the process of natural language understanding for the recently developed NLIDBs is summarized.
We provide three timelines describing the research on rule-based, machine learning-based, and hybrid approaches, as shown in Figure \ref{fig5}.

\subsubsection{Rule-based methods}

\begin{table}[!t]
	\centering
	\caption{Rules for parsing natural language queries}
	\label{tab-rules}
	\begin{tabular}{lp{0.72\textwidth}}
		\toprule
		\textbf{Rules} & \textbf{Typical NLIDBs} \\
		\midrule
		Parse tree & PRECISE \cite{PopescuEK03}, NaLIX \cite{LiYJ05,NALIX}, Querix \cite{QUERIX}, DaNaLIX \cite{LiCYSJ07}, gAnswer \cite{HuangZ13}, NaLIR \cite{LiJ14SIGMOD,LiJ14VLDB,LiJ16}, NL2TRANQUYL \cite{BoothECW15}, Unnamed method \cite{JosephPM16}, MyNLIDB \cite{DasB29} \\
		Ontology & QuestIO \cite{DamljanovicTB08}, ATHENA \cite{SahaFSMMO16}, FINESSE \cite{JammiSMVPALKSS18}, Unnamed method \cite{ErekhinskayaSPB20}, CNL-RDF-Query \cite{Henarejos-Blasco20}, ATHENA++ \cite{athena++}, Unnamed method \cite{Al-MuhammedL22} \\
		Semantic graph & Unnamed method \cite{ZouHWYHZ14}, MEANS \cite{AbachaZ15}, NL2CM \cite{AmsterdamerKM15SIGMOD,AmsterdamerKM15VLDB} \\ 
		Template matching & SQLizer \cite{Yaghmazadeh0DD17}, Unnamed method \cite{AhkoukM20}, LogicalBeam \cite{BhaskarTSS23} \\
		Pattern matching & SODA \cite{BlunschiJKMS12} \\
		Context-free grammar & TR Discover \cite{SongSSBZBMDDMH15} \\
		Semantic grammar & Unnamed method \cite{FulfordO17} \\
		\bottomrule
	\end{tabular}
\end{table}

The semantic parsing of mature NLIs is predominantly based on rules. 
The systems require specific rules to parse natural language queries, including parse tree, ontology, semantic graph, template matching, pattern matching, context-free grammar, and semantic grammar, as shown in Table \ref{tab-rules}.
Rule-based systems can only deal with knowledge bases in fixed domains and are generally not portable to other knowledge bases.
In order to enhance the accuracy of semantic understanding, systems are typically constrained by limitations in their ability to support natural language features, such as grammar and vocabulary \cite{WangX021}. 
PRECISE \cite{PopescuEK03} elucidates the notion of semantic tractability and delineates a specific subset of natural language that can be accurately converted into SQL.
However, natural language queries that cannot be processed semantically will be rejected by PRECISE.
NaLIX \cite{LiYJ05,NALIX} restricts natural language queries to a regulated subset according to a predetermined grammar.
DaNaLIX \cite{LiCYSJ07} is constructed on NaLIX and employs domain knowledge for query translation.
Domain knowledge is encapsulated within a collection of regulations that map terms with domain meaning in the parse tree to terms that can be understood by a generic system such as NaLIX.
The domain adapter within DaNaLIX assesses the current domain expertise and modifies the parse tree with related rules.
NaLIR \cite{LiJ14SIGMOD,LiJ14VLDB,LiJ16} identifies nodes within the language parse tree that have the potential to correspond to SQL components resulting from the preprocessing step, and represents semantic coverage as a subset of the parse tree. 
Such a tree explicitly corresponds to SQL and serves as a query tree, which mediates between NLQ and SQL. 
To comprehend the challenge of integrating individual and collective knowledge, NL2CM first uses RDF to represent individual and general knowledge.
Individual expression detectors are then used to distinguish between individual and general query components, which are created through a declarative selection schema in conjunction with a specialized vocabulary.
ATHENA uses domain-specific ontology to transform the natural language input into an intermediate language on the ontology. 
The intermediate language is then used to describe the semantic entities in the domain, as well as the relationships between the entities.
Ontology provides richer semantic information than relational schema, including inheritance and membership. 
By reasoning about the ontology, ATHENA demonstrates the capability to effectively discern and capture the intentions of users.
However, ATHENA is highly sensitive to changes and interpretations of user queries \cite{OzcanQSLE20}.
Both the NLIDB system described in the paper \cite{SenOQSMJLSS19} and ATHENA++ \cite{athena++} are extensions of ATHENA.
They combine linguistic analysis with deep domain reasoning to translate complex join and nested SQL.
NL2TRANQUYL \cite{BoothECW15} is a system designed for the planning of journeys within a complex multi-modal transportation system, taking into account a number of constraints, including the minimization of journey time, distance and cost. NL2TRANQUYL utilizes the ontology comprising a range of concepts to store and model related information, and generates knowledge graphs to determine the relationships between them. 
To discover and process temporal information in NLQ, TEQUILA decomposes the detected temporal problems and rewrites the generated sub-problems. 
These papers \cite{HuangZ13,ZouHWYHZ14} utilize the Stanford Parser to generate dependency trees and extract semantic relations from the parsed data. 
Subsequently, a semantic query graph is constructed by connecting these semantic relations to depict the user's query intent. 
Querix \cite{QUERIX} examines the syntax of natural language using a syntactic analyzer, which is only effective when the natural language components are complete. 
Incomplete components may result in inaccurate results, which could compromise the accuracy of the final results.

An optimal NLIDB enables users to formulate intricate queries on the database system and retrieve precise information with minimal exertion. 
Consequently, a number of systems incorporate user interaction during the process of comprehending semantics. 
NaLIX and DialSQL \cite{YavuzGSY18} adjust the query during following user engagements to revise the parse tree, however, the revision frequently necessitates a high number of user interactions.
DaNaLIX acquires domain knowledge through the interaction that occurs between the user and the system in an automated manner.
In addition to elucidating the user on the query processing procedure, NaLIR also presents a spectrum of interpretations for the user to select from, thus alleviating the user's need to address potential misunderstandings.
NaLIR is capable of detecting the parse tree, thereby enabling users to modify the parse tree directly, rather than reformulating the natural language query.
NaLIR can provide recommendations to users for revising their queries in instances where the natural language queries fall beyond the semantic boundaries.
QUICK \cite{ZenzZMSN09} improves user interactions by utilizing keyword search to enrich the expressiveness of semantic queries. 
In practical application, QUICK assists users in determining the specific intent behind natural language through a series of iterative refinement steps following the initial submission of a keyword-based question.
NLQ/A \cite{ZhengC0YZ17} enhances the user interaction component in order to more effectively address the issue of ambiguity.
SPARKLIS \cite{Ferre17} employs a sequential process consisting of three stages in order to guarantee the thoroughness of user input during searches for concepts, entities or modifiers.
While interacting with the system may result in the user feeling constrained, slowed down, and less natural when entering a query, SPARKLIS provides guidance and safety through intermediate answers and suggestions \cite{AffolterSB19}. 
In order to reduce user involvement during the disambiguation process, ATHENA utilizes the extensive semantic data within the ontology to produce a prioritized list of explanations, and employs a ranking algorithm that is intuitive and relies on ontology metrics to determine the most appropriate explanation.

\subsubsection{Machine learning-based methods}

As the usage of statistical learning methods continues to expand, there has been a growing interest in conducting semantic analysis on sentences through a variety of forms of supervision.
\textit{Pasupat and Liang} \cite{PasupatL15} employ question-and-answer format to provide guidance in responding to intricate natural language queries presented within semi-structured tables.
The paper \cite{PoonD09} represents the inaugural attempt to develop a semantic parsing model through unsupervised learning \cite{KamathD19}. 
\textit{Artzi and Zettlemoyer} \cite{ArtziZ11} solicit feedback during the conversation to determine the meaning of the user's statements.
In a domain where no training examples are available, \textit{Wang et al.} \cite{WangBL15} demonstrate the successful development of a semantic parser.
Their approach comprises two key elements: (i) \textit{a builder} and (ii) \textit{a domain-general grammar}. 
\textit{Wong and Mooney} \cite{WongM06} utilize statistical machine translation technology for the purpose of accomplishing semantic parsing tasks.

\begin{table}[!t]
	\centering
	\caption{NLIDBs with encoder-decoder frameworks}
	\label{tab-ML}
	\begin{tabular}{p{0.24\textwidth}lp{0.31\textwidth}p{0.29\textwidth}}
		\toprule
		\textbf{NLIDB} & \textbf{Year} & \textbf{Encoder} & \textbf{Decoder} \\
		\midrule
		DialSQL \cite{YavuzGSY18} & 2018 & Encode dialogue history using RNN networks & Decode errors and candidate selections \\
		SyntaxSQLNet \cite{YuYYZWLR18} & 2018 & Table-aware column encoder & Syntax tree-based decoder \\
		Unnamed method \cite{LiuCCSZ20} & 2020 & Encode NLQs and table headers using XLNet \cite{YangDYCSL19} & The parsing layer splices the vector \\
		ValueNet \cite{BrunnerS21} & 2021 & Extension of IRNet's encoder & LSTM architecture and multiple pointer networks \\
		Unnamed method \cite{DeguA21} & 2021 & The encoder of LSTM & The decoder of LSTM \\
		MIE \cite{WangLZH21} & 2021 & Multi-integrated encoder with three integrated modules & No decoder \\
		Auto-Query \cite{ParikhCJHSBA22} & 2022 & The encoder of RATSQL \cite{WangSLPR20} & SmBoP \cite{RubinB21} \\
		STAMP \cite{GiaquintoZKLTBN23} & 2023 & The encoder of T5 & The decoder of T5 \\
		Unnamed method \cite{YangZY23} & 2023 & The encoder of Transformer & The decoder of Transformer \\
		\bottomrule
	\end{tabular}
\end{table}

In recent times, there has been a growing utilization of encoder-decoder frameworks that rely on recurrent neural networks for semantic parsing, as demonstrated in Table \ref{tab-ML}.
Many systems combine machine learning and deterministic algorithms to generate structured languages \cite{word2vec}. 
This method allows the direct acquisition of the correlation between natural language and the semantic representation, eliminating the need for an intermediate representation like a parse tree \cite{KamathD19}.
Mapping natural language directly to the semantic representation can reduce the dependence of rule-based semantic parsing models on preset vocabulary, templates, and hand-generated features.
Machine learning-based models are not limited to specific knowledge bases or logical formal expressions, thus enabling the implementation of natural language interfaces that support cross-knowledge bases or cross-languages.
\textit{Wang et al.} \cite{Wang19,WangTWK20} propose a cross-domain NLI, which translates the marked natural language into the intermediate representation of the target query type by building a cross-domain multilingual \underline{seq}uence-\underline{to}-\underline{seq}uence (seq2seq) model. 
Symbols inserted into the natural language query are utilized to substitute the data elements present in the intermediate query.
However, this method is a supervised machine learning model whose effectiveness is closely related to the quality of the training data.
To ensure the accuracy of semantic understanding, a substantial quantity of training data must be provided to the model. 
A number of researchers employ a synthetic data generator as a solution to the challenge of having a restricted amount of training data available. 
The paper \cite{YuYYZWLR18} introduces SyntaxSQLNet, which can generate NLQ data sets for cross-domain SQL single-table operations, solely as a means of augmenting the training set.
The method outlined in the paper \cite{WeirU19} encompasses single-table and multi-table join queries of SQL, and can be utilized as either an augmentation or as a standalone training data set.
In terms of model training, the rule-based method is more effective than the neural network-based method, which requires more training parameters and takes longer to establish the model, consuming more memory space.

One of the earliest examples of machine learning-based systems is demonstrated in the paper \cite{ZelleM96}.
This work utilizes a deterministic shift-reduce parser and develops a learning algorithm called CHILL to learn the governing rules of parsing on the basis of inductive logic programming techniques. The corpus is trained using the CHILL method to build the parser. 
Instead of learning dictionaries, this approach assumes that a dictionary is created in advance that pairs words with semantic content rather than grammar.
The paper \cite{ZettlemoyerC05} translates the meaning of natural language sentences into lambda calculus encoding. 
The paper \cite{ZettlemoyerC05} outlines a learning algorithm whose input is a collection of sentences identified as lambda calculus expressions, and applies the method to the task of learning NLIDB to build a parser. 
While providing considerable flexibility, encoder-decoder frameworks frequently lack the ability to interpret and understand combinations of meaning \cite{KamathD19}. 
The method employed by \textit{Cheng et al.} \cite{ChengRSL17} involves the construction of the intermediate structure in two stages, which facilitates a comprehensive understanding of the model's learning process.
Similarly, the paper \cite{LapataD18} also produces an intermediate template that presents the final output in a preliminary format, thereby facilitating the subsequent decoding process.
\textit{Yin and Neubig} \cite{YinN17} address the issue of insufficient training data by incorporating explicit constraints for decoders through the utilization of target language syntax.
The approach enables the model to concentrate on parsing, directed by established grammar rules.
\textit{Xiao et al.} \cite{XiaoDG16} utilize the grammar model as prior knowledge, requiring the creation of a derivation tree while adhering to the constraints imposed by the grammar.
The approach in the paper \cite{KrishnamurthyDG17} can significantly outperform the Seq2Tree model from the aforementioned paper \cite{DongL16} by verifying that the decoder's forecasts adhere to the type constraints outlined in the type constraint grammar.
This suggests that satisfying type constraints and good formatting are equally important when generating logical expressions. 
SpatialNLI \cite{SpatialNLI,WangLKW24} is a natural language interface for the spatial field that employs the seq2seq model to understand the semantic structure of natural language, while utilizing an external spatial understanding model to identify the meaning of spatial entities. 
Subsequently, the spatial semantics learned from the spatial understanding model are integrated into natural language problems, thereby reducing the necessity of acquiring specific spatial semantics. 
SpatialNLI represents a pioneering system that integrates an external spatial semantic comprehension model to optimize the effectiveness of the principal seq2seq model.
The paper \cite{TongZY19} uses a tree model to analyze the target entity in natural language, and employs a tree-structured LSTM to understand the problem. 
The paper \cite{IyerKCKZ17} adjusts the neural sequence model to directly convert natural language into SQL, thus circumventing the intermediate query language representation. 
Then the user feedback is utilized to mark error queries, which are directly used to improve the model. 
The complete feedback loop does not necessitate the use of any intermediate language representation and is not limited to a specific domain. 
This method offers the benefit of enabling the rapid and straightforward construction of a semantic parser from scratch, and the performance of the parser improves as user feedback increases.
The encoder of ValueNet \cite{BrunnerS21} is an extension of the encoder of IRNet \cite{GuoZGXLLZ19}, receiving not only details regarding the database schema, but also extracted value candidates from the database content.

\subsubsection{Hybrid methods based on rule and machine learning}

Hybrid methods integrate rules and machine learning techniques to capitalize on the respective strengths of each, thereby enhancing the ability of the system to understand and process NLQs \cite{Koutrika24}.
Table \ref{tab-Hybrid} enumerates the representative systems that employ the hybrid approach.
Hybrid approaches are highly flexible and adaptable, as they can utilize rules for tasks with explicit rules as well as machine learning models for complex and ambiguous semantic tasks.
In addition, hybrid methods can flexibly incorporate new rules or train new machine learning models as needed to accommodate the requirements of diverse domains and tasks, and are highly scalable \cite{VisperasABACP23}.
TypeSQL \cite{YuLZZR18}, like SQLNet \cite{abs-1711-04436}, is built on sketches and formats translation tasks as slot-filling problems. The difference is that TypeSQL employs type information to enhance the understanding of entities and numbers in NLQs.
TypeSQL assigns a type to each word, such as entity, column, number and date, within the knowledge graph.
Subsequently, two bidirectional LSTM networks are utilized to encode the words in the NLQ with the corresponding column names and types.
Finally, the LSTM output hidden states are leveraged to forecast the slot values within the SQL sketch.
NALMO is a natural language interface for moving objects. To understand NLQs, NALMO employs an entity extraction algorithm to obtain entity information, including time, location and the number of nearest neighbors.
A pre-constructed corpus is then trained using LSTM to determine the query type. 
Veezoo \cite{LehmannGHSMS22} uses a range of techniques, including temporal expression parsing, entity linking, and relation extraction, to identify key information in NLQs.
The information is then extended and combined using predefined rules to generate multiple candidate intermediate representations.
Finally, Veezoo utilizes a machine learning model to score these intermediate representations in order to select the most probable interpretation of the NLQ. 
The process of data preparation in GAR \cite{FanHRGCZCJZW23} commences with a collection of sample SQLs that are tailored to a specific database.
For a given NLQ, GAR searches for the NLQs generated during data preparation and employs a learning-to-rank model to identify the most relevant query, which is then used to obtain the translation result.
The learning-to-rank model learns to rank the semantic similarities from NLQs to generated NLQs and then finds the best matching expression for a given NLQ. 
GENSQL \cite{FanRHWZL23}, a generative NLIDB, utilizes a given example SQL from the database (e.g., from query logs) to comprehend the unique structure and semantics of a given database, thereby guaranteeing precise translation outcomes.
The fundamental model used in GENSQL for converting natural language to SQL is GAR.
CatSQL \cite{FuLWLTS23} is a method for the generation of SQL that makes use of sketches.
In addition, semantic constraints are merged into the neural network-driven SQL generation procedure for semantic refinement.
CatSQL sketches are templates with keywords and slots.
CatSQL employs a deep learning algorithm to populate vacant slots in order to generate the ultimate SQL.
The deep learning algorithm is developed to focus on the generation of essential NLQ-related information, with the objective of filling the gaps without requiring the explicit generation of keywords like SELECT, FROM, and WHERE.

\begin{table}[!t]
	\centering
	\caption{NLIDBs based on rules and machine learning techniques}
	\label{tab-Hybrid}
	\begin{tabular}{p{0.165\textwidth}p{0.04\textwidth}p{0.34\textwidth}p{0.36\textwidth}}
		\toprule
		\textbf{NLIDB} & \textbf{Year} & \textbf{Rule} & \textbf{Machine learning technique} \\
		\midrule
		Unnamed method \cite{GiordaniM12} & 2012 & Generate candidate SQLs via rules and heuristic weighting schemes & Reorder candidate SQLs using the SVM sorter \\
		TypeSQL \cite{YuLZZR18} & 2018 & Assign a type to each word to understand the entity & Encode using bidirectional LSTM \\
		NALMO \cite{WangX021,WangLXL23} & 2021 & Semantic grammar and template matching & Identify the query type using LSTM \\
		Veezoo \cite{LehmannGHSMS22} & 2022 & Knowledge graph & Score intermediate representations using machine learning models \\
		GAR \cite{FanHRGCZCJZW23} & 2023 & Parse tree & Find the matching expression for NLQ using a learn-to-rank model \\
		GENSQL \cite{FanRHWZL23} & 2023 & Capture the structure of the database with sample SQLs & Find the matching expression for NLQ using a learn-to-rank model \\
		CatSQL \cite{FuLWLTS23} & 2023 & Template matching & The decoder of Transformer. Train the model using Adam \cite{KingmaB14} \\
		NALSpatial \cite{LiuWX023} & 2023 & Semantic grammar and template matching  & Identify the query type using LSTM \\
		NALSD \cite{LiuWX23} & 2023 & Semantic grammar and template matching  & Identify the query type using LSTM \\
		xDBTagger \cite{UstaKU24} & 2024 & Semantic graph & Bidirectional recurrent neural network \\
		\bottomrule
	\end{tabular}
\end{table}

Hybrid approaches based on rules and machine learning offer several advantages, including flexibility, accuracy, and scalability.
Nevertheless, such approaches present certain challenges, such as complexity, dependence on data, and tuning difficulties \cite{KatsogiannisMeimarakisK23}.
Hybrid methods require the simultaneous management and maintenance of rule engines and machine learning models, including rule definition, feature engineering, and model training, and thus have high complexity \cite{GiordaniM12}.
Furthermore, the rules and machine learning models utilized in hybrid approaches may encounter parameter tuning problems, which necessitate a significant investment of time and effort for optimization and debugging, thus increasing the costs associated with the development and maintenance of the system.
During the design and implementation of natural language interfaces, it is essential to take a comprehensive view of the advantages and challenges involved, and to make trade-offs and choices in accordance with the specific needs.

\subsection{Natural language translation}

The natural language translation stage employs the semantic information derived from the natural language understanding stage, subsequently integrating the underlying structure of the database to transform the input natural language into the corresponding executable language. 
A prevalent approach for translation is to employ complex algorithms and machine learning models to generate structured language based on the domain knowledge of the underlying database and the semantic representation of natural language \cite{2018Li}. 
Most established NLIDBs construct queries by query combination, mapping key information expressed in natural languages to corresponding components in structured languages.
We examine the process of natural language translation in recently developed NLIDBs, and summarize the general construction process of executable languages for relational and spatio-temporal databases.

\begin{figure}[!t]
	\centering
	\includegraphics[width=0.7\textwidth]{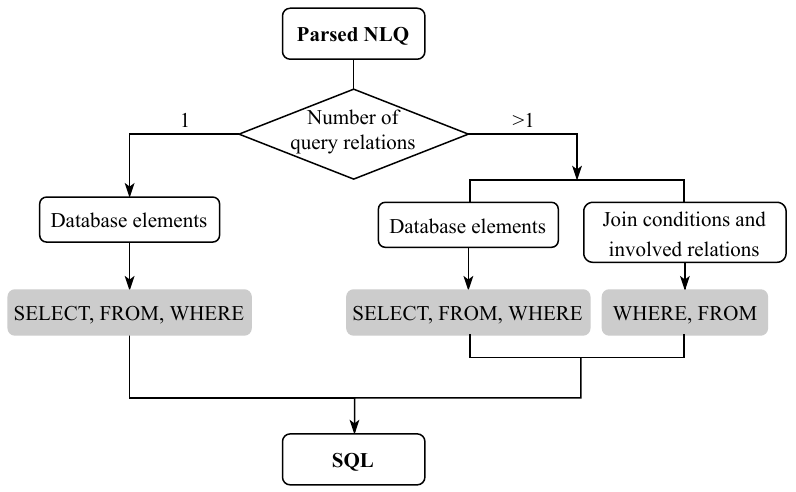}
	\caption{General build process for SQL}
	\label{fig6-1}
\end{figure}

The general build process for SQL is illustrated in Figure \ref{fig6-1} and further elaborated in the subsequent two cases.

(i) When querying a single relation, it is only necessary to place the database elements matched by NLQ in the correct positions in the SELECT, FROM, and WHERE parts, respectively. Then, SQL can be composed directly. 

(ii) When querying multiple relations, the join condition and the names of the participating relations need to be included in the WHERE and FROM clauses, respectively. 
Additionally, it is necessary to determine whether the join path is unique. 
If only one join path is available, SQL can be generated directly.
Otherwise, a query is typically generated for each possible join path, and then the most probable one is selected according to the corresponding algorithm.

\begin{figure}[!t]
	\centering
	\includegraphics[width=\textwidth]{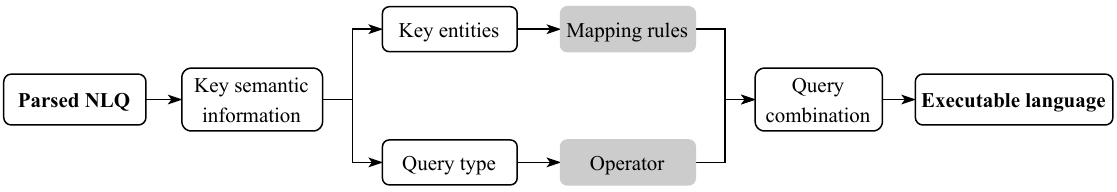}
	\caption{General construction process for executable languages over spatio-temporal data}
	\label{fig6-2}
\end{figure}

In recent years, there has been significant interest in NLI for spatio-temporal databases \cite{freya}. 
Temporal and spatial concepts are derived from the natural language description using symbolic representations in order to depict spatio-temporal features and their relationships \cite{BishopHAR18,PopescuAEKY04}. 
Due to the particularity and expressiveness of spatio-temporal problems, executable query languages over spatio-temporal databases are quite different from SQL.
Consequently, the method employed for the construction of SQL cannot be directly applied to the generation of executable languages over spatio-temporal databases.
The general process for the construction of an executable language for a spatio-temporal database is shown in Figure \ref{fig6-2}. 
Preliminary parsing of the input natural language query is performed to obtain semantic information including key entities and query types.
Subsequently, the operators necessary to construct the executable language are determined according to the type of query.
Finally, key entities and operators are combined according to certain rules to compose an executable database language.
Taking the range query over spatial data as an example, the key entities involved include spatial relations and locations.
The operator \textit{intersects} will return all objects in the relation that intersect the location if the spatial attribute of the relation and the data type of the location are both \textit{line} or \textit{region}. 
Conversely, the operator \textit{intersects} will return all objects in the relation that lie within the location if the spatial attribute of the relation is \textit{point} and the data type of the location is \textit{region}.

%First, the query type of the input NLQ is determined. 
%Operators needed to build the executable query are then determined based on the query type.
%Finally, combined with the key semantic information obtained from the semantic understanding part, a complete executable query statement can be formed.

Different DBMSs and structured languages offer a range of clauses and operators for various queries. 
%The mapping values required for translation are also distinct.
ATHENA employs a mapping strategy that correlates the ontology with the database schema in order to convert the intermediate query language utilized in the ontology into SQL.
The system described in the paper \cite{JammiSMVPALKSS18} extends ATHENA to access multiple structured backends, which is achieved through the automated translation of the intermediate query language into the specific structured query language utilized by these backend stores.
NLPQC \cite{NLPQC} is capable of processing queries formulated using predefined domain-specific templates.
Querix selectively isolates specific elements from the syntactic tree in order to align acquired knowledge with the knowledge base, thereby obtaining the final outcome. 
NL2CM leverages crowd intelligence by converting audience queries into OASSIS-QL (an extended version of SPARQL).
NaLIR utilizes the structure of the user-validated query tree to produce the suitable structure in the SQL statement and determine the join path. 
In order to ascertain whether the target SQL contains aggregate functions or sub-queries, NaLIR initially identifies function nodes or quantifier nodes in the query tree and subsequently generates SQL statements based on the identified conditions.
The paper \cite{GiordaniM12} employs lexical dependencies found in the question and database metadata to build a reasonable collection of SELECT, WHERE, and FROM clauses that enhance the quality of meaningful joins.
The paper \cite{GiordaniM12} combines clauses through a rule and heuristic weighting scheme, and then generates a sorted list of candidate SQLs, demonstrating that full semantic interpretation can be avoided by relying on a simple SQL generator. 
This method can be employed iteratively to address intricate issues necessitating nested SELECT commands.
Finally, this paper \cite{GiordaniM12} applies the re-ranker to reorder the list of questions and SQL candidate pairs with the aim of enhancing the accuracy of the system.
TEQUILA uses a standard KB-QA system to evaluate the sub-questions from the semantic understanding part individually. The results of the sub-questions are then combined with the reasoning to calculate the answer to the full question.
NL2TRANQUYL translates English requests into formal TRANQUYL \cite{BoothSWC09} queries using the knowledge graph generated by the semantic comprehension component. 
The traffic query language TRANQUYL for travel planning follows the conventional SQL structure of ``\textit{SELECT, FROM, WHERE}".
NALMO supports five distinct types of moving object queries, including (i) \textit{time interval queries}, (ii) \textit{range queries}, (iii) \textit{nearest neighbor queries}, (iv) \textit{trajectory similarity queries}, and (v) \textit{join queries}. 
In the query translation process, NALMO first constructs a corpus comprising the five query types, collectively referred to as MOQ. 
Then the LSTM neural network is used for training, resulting in a model that is capable of accurately identifying the specific type of query. Finally, the appropriate operators are selected according to the query type, and the entity information extracted by the semantic parsing component is combined to build the executable language for SECONDO.

\section{NL2SQL benchmarks}\label{sec4}

We presents 11 frequently used benchmarks for transforming NLQ into SQL and three evaluation metrics, exploring the methods for generating new benchmarks.

\subsection{Existing benchmarks}

The details of NLQ and executable language pairs for common domains are presented in Table \ref{tab4}. 
The majority of existing benchmarks are utilized in the domain of relational databases to transform \underline{n}atural \underline{l}anguage query in\underline{to} \underline{SQL} (NL2SQL).
The comparison of fields and types of SQL supported by the benchmarks for NL2SQL is shown in Table \ref{tab5}.
We can conclude that GeoQuery and Spider support the most types of SQL, while WikiSQL supports only the simple select query.
The queries in WikiSQL and Spider cover a multitude of domains.
In recent years, GeoQuery, MAS, WikiSQL and Spider have been employed with considerable frequency.

\begin{table}[!t]
	\caption{Examples of NLQ and executable language pairs for common domains}
	\label{tab4}
	\begin{tabular}{p{0.2\textwidth}p{0.75\textwidth}}
		\toprule
		\textbf{Domain} & \textbf{Examples of NLQ and executable language pairs} \\
		\midrule
		\multirow{3}{*}{Relational database} & 
		\textbf{NLQ:} How many CFL teams are from York College?\\
		& \textbf{SQL:}\\
		& SELECT COUNT CFL Team FROM CFLDraft WHERE College = ``York"\\
		\midrule
		\multirow{6}{*}{Spatial domain} & \textbf{NLQ1:} What is the population of San Antonio?\\
		& \textbf{Lambda expression:}\\
		& answer(A,population(B,A),const(B,cityid(San Antonio)))\\
		& \textbf{NLQ2:} Could you tell me what parks are in the center?\\
		& \textbf{Executable language:}\\
		& query park feed filter [.GeoData ininterior center] consume;\\
		\midrule
		\multirow{5}{*}{Moving Objects} & 
		\textbf{NLQ:} Where did the train 7 go at 8am?\\
		& \textbf{Executable language:}\\
		& query Trains feed filter [.Id = 7] filter [.Trip present [const instant value ``2020-11-20-8:00"]] extend [Pos: val (.Trip atinstant [const instant value ``2020-11-20-8:00"])] project [Id, Line, Pos] consume;\\
		\midrule
		\multirow{5}{*}{Trip planning} & 
		\textbf{NLQ:} Can I walk to 300 W. Humboldt Blvd. by 4:00 p.m.?\\
		& \textbf{TRANQUYL:} \\
		& SELECT $ \ast $ FROM ALL\_TRIPS(user.current\_location, 300 W. Humboldt Blvd.) AS t WITH MODES pedestrian WITH CERTAINTY .78 WHERE ENDS(t) $\leq$ 4:00 p.m. MINIMIZE DURATION(t)\\
		\midrule
		\multirow{7}{*}{Crowd mining} & 
		\textbf{NLQ:} What are the most interesting places near Forest Hotel, Buffalo, we should visit in the fall?\\
		& \textbf{OASSIS-QL:}\\
		& SELECT VARIABLES \$x WHERE \{\$x instanceOf Place. \$x near Forest\_Hotel,\_Buffalo,\_NY\} SATISFYING \{\$x hasLabel ``interesting"\} ORDER BY DESC(SUPPORT) LIMIT 5 AND \{[] visit \$x. [] in Fall\} WITH SUPPORT THRESHOLD = 0.1\\
		\bottomrule
	\end{tabular}
\end{table}

\begin{table}[!t]
	\centering
	\caption{Comparison of benchmarks for NL2SQL}
	\label{tab5}
	\begin{tabular}{p{0.17\textwidth}|c|c|c|c|c|p{0.245\textwidth}|p{0.26\textwidth}}
		\hline
		\textbf{Benchmark} & \rotatebox{90}{\textbf{Select query}} & \rotatebox{90}{\textbf{Group query}} & \rotatebox{90}{\textbf{Sort query}} & \rotatebox{90}{\textbf{Join query}} & \rotatebox{90}{\textbf{Nested query}} & \textbf{Fields involved} & \textbf{Usage in papers} \\
		\hline
		ATIS & \checkmark &  &  & \checkmark & \checkmark & air travel & \cite{IyerKCKZ17,RadevKZZFRS18,PazosFBO20,PazosFL20,SidorovRFCL20} \\
		\hline
		Restaurant & \checkmark & & & \checkmark & & restaurant & \cite{TangM00,PopescuEK03,LiCYSJ07,SpatialNLI} \\
		\hline
		GeoQuery & \checkmark & \checkmark & \checkmark & \checkmark & \checkmark & geography & \cite{TangM00,TangM01,PopescuEK03,ZettlemoyerC05,LiCYSJ07,GiordaniM12,SahaFSMMO16,IyerKCKZ17,RadevKZZFRS18,SpatialNLI,PazosFL20,athena++,SidorovRFCL20,FanHRGCZCJZW23,WangLKW24} \\
		\hline
		MAS & \checkmark & \checkmark & & \checkmark & & academic & \cite{LiJ14VLDB,SahaFSMMO16,Yaghmazadeh0DD17,DeutchFGH18,BaikJ019,athena++,UstaKU21,UstaKU24} \\
		\hline
		Scholar & \checkmark & & & \checkmark & & academic & \cite{RadevKZZFRS18} \\
		\hline
		IMDB & \checkmark & & & \checkmark & &  internet movie & \cite{Yaghmazadeh0DD17,BaikJ019,Henarejos-Blasco20,UstaKU21,UstaKU24} \\
		\hline
		YELP & \checkmark & & & \checkmark & & business review & \cite{Yaghmazadeh0DD17,BaikJ019,UstaKU21,UstaKU24} \\
		\hline
		WikiSQL & \checkmark & & & & & 
		multiple fields (e.g. state, college, manufacturer) & \cite{abs-1709-00103,YavuzGSY18,YuLZZR18,YaoSSY19,LiuCCSZ20,WangTWK20,FuLWLTS23,GiaquintoZKLTBN23,SunGZSCLS23} \\
		\hline
		ParaphraseBench & \checkmark & \checkmark & & & & medical & \cite{abs-1804-00401} \\
		\hline
		Advising & \checkmark & & & \checkmark & \checkmark & university course & \cite{RadevKZZFRS18} \\
		\hline
		Spider & \checkmark & \checkmark & \checkmark & \checkmark & \checkmark & 
		138 different fields (e.g. car, stadium, country)
		& \cite{YuZYYWLMLYRZR18,GuoZGXLLZ19,YaoSSY19,athena++,BrunnerS21,GanCXPWDZ21,MellahREBB21,UstaKU21,FanHRGCZCJZW23,FanRGZHWWS23,FuLWLTS23,GiaquintoZKLTBN23,SunGZSCLS23} \\
		
		\hline
	\end{tabular}
\end{table}

The details of popular benchmarks are shown in Table \ref{tab6}. Early data sets consist of only one domain and one database, such as ATIS, Restaurant and GeoQuery. 
In contrast, the latest data sets, for example WikiSQL and Spider, contain multiple domains and several databases with larger and more diverse NLQs and SQLs.

\begin{table}[!t]
	\centering
	\caption{Details of popular benchmarks}
	\label{tab6}
	\begin{tabular}{lcccc}
		\toprule
		\textbf{Benchmark} & \textbf{Year} & \textbf{\#queries} & \textbf{\#tables} & \textbf{Domains covered} \\
		\midrule
		ATIS \cite{Price90}      & 1990 & 5871  & 25    & single field \\
		Restaurant \cite{TangM00} & 2000 & 250 & 3 & single field \\
		GeoQuery \cite{TangM01}  & 2001 & 880   & 7     & single field \\
		MAS \cite{LiJ14VLDB} & 2014 & 196 & 17 & single field \\
		Scholar \cite{IyerKCKZ17}   & 2017 & 816 & 10 & single field \\
		IMDB \cite{Yaghmazadeh0DD17}      & 2017 & 131 & 16 & single field \\
		YELP \cite{Yaghmazadeh0DD17}      & 2017 & 128 & 7 & single field \\
		WikiSQL \cite{abs-1709-00103}    & 2017 & 80654 & 24241 & multiple fields \\
		ParaphraseBench \cite{abs-1804-00401}  & 2018 & 290 & 1 & single field \\ 
		Advising \cite{RadevKZZFRS18}  & 2018 & 4387  & 15    & single field \\
		Spider \cite{YuZYYWLMLYRZR18}    & 2018 & 10181 & 1020 & multiple fields \\
		\bottomrule
	\end{tabular}
\end{table}

(i) \textbf{ATIS} (\underline{A}irline \underline{T}ravel \underline{I}nformation \underline{S}ystem) \cite{Price90} is a classical data set with a relatively old age, having been introduced by Texas Instruments in 1990.
ATIS is built on the relational database Official Airline Guide, comprising 25 tables and 5871 queries written in English.
The queries pertain to details regarding flights, ticket prices, destinations, and services available at airports.
The queries in ATIS are for the air travel field, including join queries and nested queries, but no grouping and sorting queries. 
The average length of NLQs and SQLs in ATIS is approximately 11 and 67 words, respectively.
Each query operates on an average of six tables. An example query is as follows.

%\vspace{0.5em}
\begin{flushleft}
	\textit{\textbf{Q1:} What aircraft is used on delta flight 1984 from Kansas city to Salt Lake city?}
\end{flushleft}
%\vspace{0.5em}

(ii) \textbf{Restaurant} \cite{TangM00} comprises a vast collection of dining establishments located in Northern California, storing restaurant names, locations, features, and travel guide ratings. 
The benchmark contains 250 questions about restaurants, food types and locations. An example query is as follows.

%\vspace{0.5em}
\begin{flushleft}
	\textit{\textbf{Q2:} Where is a good Chinese restaurant in Palo Alto?}
\end{flushleft}
%\vspace{0.5em}

(iii) \textbf{GeoQuery} \cite{TangM01} consists of 8 tables and 880 natural language queries in the US geographic database. 
The queries in GeoQuery are designed for the geographic domain, including join queries, nested queries, grouping queries and sorting queries.
The average length of NLQs and SQLs in GeoQuery is about 8 and 16 words, respectively.
Additionally, each query operates on an average of one table. 
Although the queries are relatively brief in length, they are highly composable, with nearly half of the SQL containing at least one nested sub-query. One of the English queries is as follows.

%\vspace{0.5em}
\begin{flushleft}
	\textit{\textbf{Q3:} What is the largest city in states that border California?}
\end{flushleft}
%\vspace{0.5em}

(iv) \textbf{MAS} \cite{LiJ14VLDB} is generated from the Microsoft Academic Search database, which stores information such as academic papers, authors, journals, and conferences. 
The source of NLQs in MAS is the logical queries that are capable of being articulated in the search interface of the Microsoft Academic Search platform. 
The fields of MAS and Scholar are both academic in nature, but exhibit distinct patterns. One English query is as follows.

%\vspace{0.5em}
\begin{flushleft}
	\textit{\textbf{Q4:} Return authors who have more papers than Bob in VLDB after 2000.}
\end{flushleft}
%\vspace{0.5em}

(v) \textbf{Scholar} \cite{IyerKCKZ17} consists of 816 NLQs for academic database search that are annotated with SQL.
The average length of NLQs and SQLs in Scholar is approximately 7 and 29 words, respectively.
Each query operates on an average of 3 tables. 
\textit{Iyer et al.} \cite{IyerKCKZ17} provide a database for performing these queries, which includes academic articles, journal details, author information, keywords, citations, and utilized datasets.
One of the English queries is as follows.

%\vspace{0.5em}
\begin{flushleft}
	\textit{\textbf{Q5:} Get all author having data set as DATASET\_TYPE.}
\end{flushleft}
%\vspace{0.5em}

(vi) \textbf{IMDB and YELP} \cite{Yaghmazadeh0DD17} are generated using data from the Internet Movie Database and Business Review Database, respectively. 
The NLQs are obtained from coworkers of the authors of the paper \cite{Yaghmazadeh0DD17}, who are only aware of the types of data available in the database and not the underlying database schema.

(vii) \textbf{WikiSQL} \cite{abs-1709-00103}, introduced in 2017, is a comprehensive and meticulously annotated collection of natural language to SQL mappings, and currently represents the most extensive data set for NL2SQL.
WikiSQL contains SQL table instances extracted from 24241 HTML tables on Wikipedia, and 80654 natural language queries, each accompanied by an SQL.
WikiSQL comprises genuine data extracted from the web, with queries involving a multitude of tables, but the queries do not involve complex operations such as GROUP BY and multi-table union queries.
The majority of questions in WikiSQL are between 8 and 15 words in length, most SQLs are between 8 and 11 words, and most table columns are between 5 and 7.
In addition, most natural language queries are of the \textit{what} type, followed by \textit{which}, \textit{name}, \textit{how many}, \textit{who}. 
The execution accuracy of WikiSQL has significantly improved from the initial 59.4\% to 93.0\%, and the method has undergone a transformation from a simple seq2seq approach to a multi-tasking, transfer learning, and pre-training paradigm. 
A pair of questions and SQLs for the CFLDraft table can be formulated as follows.

%\vspace{0.5em}
\begin{flushleft}
	\textit{\textbf{Q6:} How many CFL teams are from York College?}
	
	\textit{\textbf{SQL\_Q6:} SELECT COUNT CFL Team FROM CFLDraft WHERE College = ``York"}
\end{flushleft}
%\vspace{0.5em}

(viii) \textbf{ParaphraseBench} \cite{abs-1804-00401}, a component of the DBPal paper \cite{BasikHIURUWBC18}, is a benchmark utilized to assess the robustness of NLIDBs. 
Unlike existing benchmarks, ParaphraseBench covers diverse language variants of user input NLQs and maps natural language to the anticipated SQL output. 
The benchmark is constructed upon a medical database that contains a single table for storing patient information. 
The language variants utilized in NLQs permit the classification of NLQs into six categories, as illustrated in Table \ref{tab7}.

\begin{table}[!t]
	\centering
	\caption{Query categories and examples for ParaphraseBench}
	\label{tab7}
	\begin{tabular}{ll}
		\toprule
		\textbf{Category} & \textbf{Example queries} \\
		\midrule
		Naive & What is the average length of stay of patients where age is 80? \\
		Syntactic & Where age is 80, what is the average length of stay of patients? \\
		Morphological & What is the averaged length of stay of patients where age equaled 80? \\
		Lexical & What is the mean length of stay of patients where age is 80 years? \\
		Semantic & What is the average length of stay of patients older than 80? \\
		Missing Information & What is the average stay of patients who are 80? \\
		\bottomrule
	\end{tabular}
\end{table}

(ix) \textbf{Advising} \cite{RadevKZZFRS18} was proposed in 2018, and the NLQs were built on a database of course information from the University of Michigan containing fictitious student profiles.
A portion of the queries are collected from the Facebook platform of the EECS department, and the remaining questions are formulated by computer science students well-versed in database topics that might be raised in academic consulting appointments. 
The queries in Advising are for student-advising tasks, including join queries and nested queries. One of the English queries is as follows.

%\vspace{0.5em}
\begin{flushleft}
	\textit{\textbf{Q7:} For next semester, who is teaching EECS 123?}
\end{flushleft}
%\vspace{0.5em}

(x) \textbf{Spider} \cite{YuZYYWLMLYRZR18} is a large NL2SQL data set introduced by Yale University in 2018, in order to solve the requirement for extensive and high-caliber datasets for a novel intricate cross-domain semantic parsing challenge.
The data set contains 10181 natural language queries and 5693 corresponding complex SQLs, which are distributed across 200 independent databases, and the content covers 138 different domains. 
The average length of questions and SQL statements in Spider is approximately 13 and 21 words, respectively. 
While the number of questions and SQLs in Spider is not as extensive as that of WikiSQL, Spider contains all common SQL patterns and complex SQL usages, including advanced operations like HAVING, GROUP BY, ORDER BY, table joins, and nested queries, which makes Spider closely aligned with real-world scenarios.
The following is an illustrative example of a complex problem and the corresponding SQL, which contains a nested query, a GROUP BY component, and multiple table joins.

%\vspace{0.5em}
\begin{flushleft}
	\textit{\textbf{Q8:} What are the name and budget of the departments with average instructor salary greater than the overall average?}
	
	\textit{\textbf{SQL\_Q8:} SELECT T2.name, T2.budget FROM instructor as T1 JOIN department as T2 ON T1.department id = T2.id GROUP BY T1.department id HAVING avg (T1.salary) $>$ (SELECT avg (salary) FROM instructor)}
\end{flushleft}
%\vspace{0.5em}

%In addition to the benchmarks for NL2SQL described above, the paper \cite{WangX021} proposes a benchmark for moving objects queries, named MOQ (Moving Objects Queries). 
%MOQ contains 240 moving objects queries, divided into four categories, each category and its example query are shown in Table \ref{tab5-4}. 
%These queries are extracted from more than 60 conference and journal papers in the field of moving objects and then processed manually by experts.
%
%\begin{table}[!t]
%	%\centering
%	\caption{Query categories and examples for MOQ}
%	\label{tab5-4}
%	\begin{tabular}{p{0.3\textwidth}p{0.64\textwidth}}
%		\toprule
%		Category & Example query \\
%		\midrule
%		Time Interval Query & Where were the trains at a certain time $t_{1}$? \\
%		Range Query & When did the train 123 pass underground Nanjing Station? \\
%		Nearest Neighbor Query & Show me five nearest neighbors to the train 3 from 8am to 10am. \\
%		Trajectory Similarity Query & Find trains which are similar to the train 7 at any time instance of the time period [$t_{8}$, $t_{9}$]? \\
%		\bottomrule
%	\end{tabular}
%\end{table}

\subsection{Generation of new benchmarks}

Modifying an existing NL2SQL benchmark to generate a new one is a common practice.
The following steps describe the process in detail.

(i) Researchers are required to conduct a meticulous analysis of the existing benchmarks, including an examination of the data structures, query types, and complexity.
Through the analysis, they can gain insight into the constraints of the benchmark and identify potential avenues for enhancement.

(ii) Designing a modification strategy is a critical step, which involves determining how to modify and extend the benchmark on the basis of the analysis results.
The step may include adding new queries, changing the linguistic expression of queries, and introducing complex query types.

(iii) In the process of implementing modifications, researchers are expected to execute the designed modification strategy with precision in order to ensure that the new benchmark meets the expected requirements.

(iv) Evaluating the performance is a pivotal aspect of the process. The researchers employ the modified benchmark to train and test NL2SQL models, subsequently assessing the models' performance and generalization capabilities according to the new benchmark.

Building on Spider \cite{YuZYYWLMLYRZR18}, \textit{Kaoshik et al.} \cite{KaoshikPRAJ021} propose a new NL2SQL benchmark, named ACL-SQL, containing five tables and 3100 pairs of NLQ and SQL.
By defining and annotating three types of questions on temporal aspects in Spider: (i) \textit{questions querying for temporal information}, (ii) \textit{questions querying for temporal information with grouping or ordering}, and (iii) \textit{questions with temporal conditions}, \textit{Vo et al.} \cite{VoPMS22} propose a new data set, TempQ4NLIDB, which can assist NLIDB systems based on machine learning approaches to improve their performance on temporal aspects.
To address the dearth of publicly available benchmarks on ambiguous queries, \textit{Bhaskar et al.} \cite{BhaskarTSS23} generate a new benchmark called AmbiQT by modifying Spider with a combination of synonym generation and ChatGPT-based and standard rules-based perturbation.
AmbiQT comprises in excess of 3000 examples, each of which can be interpreted as two valid SQLs due to lexical ambiguities (namely, unambiguous column and table names) or structural ambiguities (namely, the necessity of joins and the pre-computation of aggregations).

In light of the limitations of existing benchmarks, including (i) \textit{the presence of data bias or linguistic expression limitations}, and (ii) \textit{the limited coverage of domains and contexts that cannot fully represent real-world diversity}, researchers have proposed generators for Text2SQL benchmarks.
\textit{Weir et al.} \cite{WeirU19} present a synthesized data generator that synthesizes SQL patterns in the template syntax, including aggregations, simple nesting, and column joins.
Each SQL pattern is matched with numerous different \underline{n}atural \underline{l}anguage (NL) patterns, allowing for the generation of a vast number of domain-specific NLQs and SQLs.
\textit{Luo et al.} \cite{Luo00CLQ21} propose an NL2VIS synthesizer, named NL2SQL-to-NL2VIS, which is capable of generating multiple pairs of natural language and VIS from a single NL and SQL pair based on semantic joins between SQL and VIS queries.
NL2SQL-to-NL2VIS can be utilized to create NL2VIS benchmarks from established NL2SQL benchmarks.
\textit{Hu et al.} \cite{HuZJLZCLPWHZGDL23} suggest a framework for synthesizing Text2SQL benchmarks. The framework involves first synthesizing SQL and then generating NLQs. 
At the stage of synthesizing SQL, a method is suggested for column sampling based on pattern distance weighting to prevent excessive complexity in concatenation.
In the process of generating text from SQL, an intermediate representation is used to facilitate the transition from SQL to NLQ, thereby enhancing the quality of the generated NLQ.

\subsection{Evaluation metrics}

NLIDB is intended to assist users in efficiently querying and retrieving query results, and thus evaluating the response time and effectiveness of the system is essential.
Response time measures how quickly the system can process a user's natural language queries and return the relevant results.
Effectiveness measures how well the system translates natural languages into accurate and relevant executable database languages, which consists of two measures: (i) \textit{translatability} and (ii) \textit{translation precision}.
%The effectiveness of the system is to judge whether the system is capable of generating executable languages and whether users can achieve the expected results.

\begin{definition}[\textbf{Translatability}]
	Given the set $ E $ of executable languages generated by the system and the set $ N $ of input natural language queries, the translatability $ T $ is defined as follows.
	%Given the number of queries accurately translated by the system $\#executable$ and the total number of input queries $\#overall\_queries$, we have
	\begin{equation}
		T = \dfrac{|E|}{|N|} \nonumber
	\end{equation}
\end{definition}

\begin{definition}[\textbf{Translation precision}]
	Given the set $ ER $ of executable languages that meet the expected results, the set $ N $ of natural language queries entered into the system, the translation precision $ TP $ is defined as follows.
	%Given the number of executable queries that produce expected results $\#expected\_results$ and the total number of input queries $\#overall\_queries$, we have
	\begin{equation}
		TP = \dfrac{|ER|}{|N|} \nonumber
	\end{equation}
\end{definition}

Response time denotes the duration necessary for the system to transform the input natural language into the executable language of the database.
This temporal interval represents the difference between the moment when the system furnishes the translated output and the moment when the natural language is received.
Translatability is a measure of the likelihood of the system accurately translating a natural language into an executable language.
This metric is quantified as the proportion of correctly translated queries out of the total number of queries submitted to the system.
Translation precision refers to the likelihood that the final output of the translated executable language matches the expected outcome, and is quantified as the ratio of executable languages producing the desired results to the overall number of queries.

The outcomes of evaluating a system may be different depending on the benchmark used.
The size of the benchmark affects the accuracy of the semantic parsing part of the system. 
Complex queries in the benchmark can be used to assess the system's ability for generalization.
In related papers, PRECISE achieves 95.0\% translatability and translation precision on the Restaurant benchmark and 77.5\% on the GeoQuery benchmark. 
ATHENA has a translatability and translation precision of 87.2\% on the GeoQuery benchmark and 88.3\% on the MAS benchmark.
The translatability and translation precision of NALMO on the benchmark MOQ are 98.1\% and 88.1\%, respectively.

\section{System interfaces development}\label{sec5}

We categorize recently developed NLIDBs according to the technical approach and the data stored in the backend.
The methods of developing and using the system interfaces are then divided into two categories for analysis and summary:(i) \textit{used as an independent software} and (ii) \textit{used as a module of a database management system}. 
Finally, enhancements to the existing NLIDB system are presented in three aspects.

\subsection{Recently developed NLIDBs}

We are concerned with the NLIDBs, which have emerged since 2000.
There are several ways to classify NLIDBs.
\textit{Affolter et al.} \cite{AffolterSB19} divide recently developed systems into four categories.

(i) \textbf{Keyword-based systems} are represented by SODA \cite{BlunschiJKMS12}. 
The core of such a system lies in the search process, where the inverted index containing fundamental data and metadata from the database is utilized as the retrieval target.
This process involves comparison with natural language, and identification of keywords referenced in the query.
Although simple, the approach fails to identify the potential semantics that are not directly present in natural language. 
Such systems are unable to respond to aggregation queries and complex questions involving sub-queries.

(ii) \textbf{Pattern-based systems}, exemplified by NLQ/A \cite{ZhengC0YZ17} and QuestIO \cite{DamljanovicTB08}, are extensions of keyword-based systems that are capable of incorporating natural language patterns and mapping to pre-specified query sentence patterns.

(iii) \textbf{Parsing-based systems} are typified by NaLIR, a general interactive natural language interface designed for querying relational databases. 
NaLIR employs the existing natural language parser to acquire the semantic understanding of the given NLQ which is represented by a parse tree, and then converts the semantic understanding into database understanding and finally into SQL.  
Such systems incorporate a multitude of natural language processing methods, including the parsing of natural language sentences employing parse trees. 
One principal benefit of this method is the ability to map semantics into predefined SQL templates.

(iv) \textbf{Grammar-based systems} are represented by TR Discover \cite{SongSSBZBMDDMH15} and MEANS \cite{AbachaZ15}. 
The foundation of such systems consists of a predetermined set of grammar rules, which are used to constrain the questions that users can pose to the system in order to form formal NLQs that are straightforward to analyze. 
The primary advantage of this approach is that the systems are capable of providing users with guidance as they enter questions, and can respond to all questions that adhere to the established rules. 
In comparison to keyword-based, pattern-based and parsing-based systems, grammar-based systems are considered to be the most robust, despite relying significantly on predefined manual rules.

\begin{figure}[!t]
	\centering
	\includegraphics[width=\textwidth]{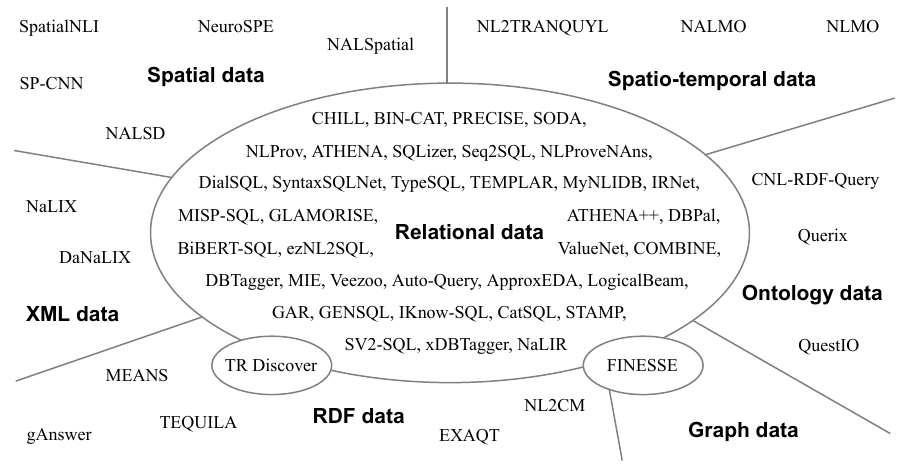}
	\caption{Classification of NLIDBs based on data stored in the backend}
	\label{fig7}
\end{figure}

In this survey, we categorize NLIDBs into seven distinct groups according to the data stored in the backend.
The representative systems for each category are depicted in Figure \ref{fig7}.
Among the various categories, natural language interfaces for relational data are the most prevalent and functional, and are subjected to ongoing research on an annual basis.
Recently, research on NLIs for XML data has not advanced, remaining at the same stage as in 2007. 
The two main reasons are (i) \textit{an increasing preference for JSON as a format for data exchange over XML}, and (ii) \textit{the suitability of NoSQL databases for handling unstructured or semi-structured data over XML databases}.
Since 2013, NLIs for natural language queries over RDF data, ontology data, graph data, spatial data, and spatio-temporal data have been developed.
The executable languages transformed by these NLIs correspond to the databases used.

NLIDB for relational data transforms natural language queries into SQL.
IRNet \cite{GuoZGXLLZ19} first identifies the entities contained in the NLQ, including columns, tables and values. 
Subsequently, a neural model based on syntax is used to synthesize an intermediate representation connecting natural language with SQL.
Finally, IRNet derives SQLs on the basis of intermediate representations.
Representative NLIDBs for XML databases are NaLIX \cite{NALIX} and DaNaLIX \cite{LiCYSJ07}, which transform natural language queries into XQuery.
NaLIX restricts natural language to a predefined subset of the grammar.
DaNaLIX builds upon NaLIX and enables users to leverage domain knowledge for query transformation.
TEQUILA \cite{JiaARSW18} is a typical NLIDB for RDF data, which transforms natural language queries into SPARQL.
%To discover and process the temporal information in the query, TEQUILA decomposes the detected temporal questions and rewrites the sub-questions.
TEQUILA employs a standard knowledge-based question and answer system to evaluate sub-questions independently. The results of the sub-questions are then combined for inference to compute the answer to the full question.
QuestIO \cite{DamljanovicTB08} works for querying structured data represented in ontology format.
Built on the ontology and a knowledge base containing instances of the ontology's concepts, QuestIO accepts NLQ as input and produces SeRQL as output.
Utilizing the language processing framework GATE, QuestIO combines fundamental concepts with keywords, blocks and phrases to deduce potential relationships among the concepts in the ontology.
In the spatio-temporal domain, NLIDB can handle GIS-related queries, such as historical meteorological data at a specific location, and geographic position information at different moments.
NeuroSPE \cite{QiuXMTZ23} is a spatial extraction model designed to identify spatial relations within Chinese natural language text.
The model extends a bidirectional gated recurrent neural network with a series of pre-trained models and is able to address specific challenges in a variety of natural language text, including the absence of direct context and the occurrence of abbreviations, special languages, and symbols.
NALMO \cite{WangX021,WangLXL23} is a natural language interface designed for moving objects that allows users to submit queries of five types, including
(i) \textit{time interval queries}, (ii) \textit{range queries}, (iii) \textit{nearest neighbor queries}, (iv) \textit{trajectory similarity queries}, and (v) \textit{join queries}.

Several systems have been created that can be used across various back-end data stores, with the objective of enhancing the generality of NLIDB.
TR Discover \cite{SongSSBZBMDDMH15} is one such system which transforms NLQ into SPARQL or SQL. 
TR Discover generates FOL representations by analyzing natural language using a feature-based context-independent grammar consisting of entries in the vocabulary for leaf nodes and rules governing the phrase structure for non-terminal nodes. 
The FOL representation is then parsed into a parse tree through the utilization of a first-order logic parser. 
The parse tree is traversed sequentially and transformed into SPARQL or SQL.
FINESSE \cite{JammiSMVPALKSS18}, an extension to ATHENA, is a system that seamlessly connects to multiple structured data stores. 
FINESSE can access various structured backends (e.g., RDF stores and Graph stores) by automatically transforming the intermediate query language OQL into the corresponding structured query language specific to the backends (e.g., SPARQL and Gremlin).

\subsection{Development and usage of system interfaces}

The combination of the aforementioned three components, including (i) \textit{natural language preprocessing}, (ii) \textit{natural language understanding} and (iii) \textit{natural language translation}, constitutes a comprehensive system architecture. 
Then the theoretical knowledge is implemented in the form of a system.
There are two primary methods of development and usage:

\begin{figure}[!t]
	\centering
	\subfigure[A stand-alone software]{
		\label{fig8-1}
		\includegraphics[scale=0.7]{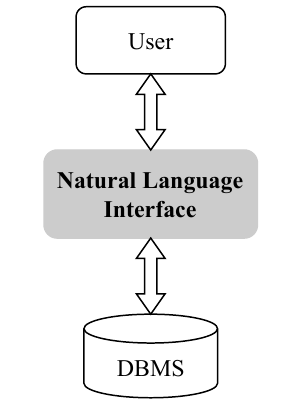}}
	\hspace{1in}
	\subfigure[A plug-in for DBMS]{
		\label{fig8-2}
		\includegraphics[scale=0.7]{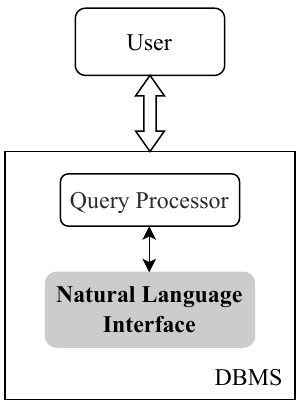}}
	\caption{The architecture of the database system with NLI}
	\label{fig8}
\end{figure}

(i) A stand-alone software. 
In this scenario, the system generally comprises a separate visual interface and a database, and the architecture is shown in Figure \ref{fig8-1}. 
A visual interface allows users to write natural language problems that are interactively translated into executable language.
By submitting the executable language in the corresponding database management system, the query results can be obtained.
The paper \cite{LiJ14SIGMOD} presents the JavaScript-driven interface of NaLIR, which interacts with a master server implemented in Java.
NL2CM is implemented in Java 7, whose web user interface is constructed in PHP 5.3 and jQuery 1.x.
The paper \cite{IyerKCKZ17} develops a web interface designed to receive NLQs from users directed towards academic databases and display translated SQLs. 
The interface also shows several example utterances to assist users in comprehending the domain. 
The tool that comes with the NLMO system is a web application written in Java.

(ii) A plug-in for the database management system. 
In this instance, the system exists in a format analogous to a Python custom module and interacts with the user through the visual interface of the database management system. 
The system architecture is illustrated in Figure \ref{fig8-2}.
The user inputs NLQ by invoking the interface provided by the system. 
Thereafter, the database management system automatically calls the NLI module to process the NLQ, and displays the translated executable language on the visual interface. 
One of the most typical systems is NALMO, which is developed on a laptop running Ubuntu 14.04.
The final interface form in SECONDO is represented as an algebraic module with an operator. 
The users can use the operator on the moving objects databases in SECONDO to perform the corresponding NLQ translation of moving objects.

%\begin{figure}[!t]
%	\centering
%	\includegraphics[width=0.65\textwidth]{fig6-2-0.pdf}
%	\caption{The architecture of the database system with NLI}
%	\label{fig6-2}
%\end{figure}

\subsection{Enhancement of NLIDB systems}

Although existing NLIDBs have been able to achieve the transformation from natural language to executable database language, the research on NLIDB is a long process and the systems need to be optimized step by step because natural language has rich expressions, ambiguous semantic knowledge and intricate correlations \cite{Ives18a}.
Enhancements to the existing NLIDB systems are mainly in the following three areas: (i) \textit{interpreting answers and non-answers to queries}, (ii) \textit{improving the effectiveness of the system}, and (iii) \textit{securing the system against potential vulnerabilities}.

\subsubsection{Interpreting answers and non-answers to queries}

Researchers have enhanced the functionalities of existing systems with regards to providing explanations for both query answers and non-answers.
Users of NLIDB do not usually have the relevant expertise and may have difficulty in understanding the results or verifying their correctness. 
In this work, papers \cite{DeutchFG16,DeutchFG17,DeutchFG18,DeutchFG20} complement these efforts by providing NL explanations for query answers. 
The authors propose a system named NLProv, which employs the original NLQ structure to transform the provenance information into natural language.
The obtained provenance information is then presented to the user in the form of natural language answers, through a four-step process: 
\begin{compactitem}
	\item The user inputs a query using natural language that is transmitted to the improved NaLIR. The system processes the NLQ, constructs a formal query, and stores the translated portions of the NLQ in relation to the formal query. 
	\item NLProv employs the SelP system \cite{DeutchGM15} to evaluate formal queries and records the provenance of each query, indicating the correlation between dependency tree nodes and specific provenance sections.
	\item The source information is decomposed and then compiled into an NL answer with explanation.
	\item The system presents the factorized answer to the user. In cases where the answer is excessively detailed and difficult to comprehend, users have the option to access summaries at various levels of nesting.
\end{compactitem}

The paper \cite{DeutchFG20} proposes a general solution for NLProv that is not specific to NaLIR. 
The core of the solution is an alternative architecture that does not depend on the query builder for producing the partial mappings between the nodes of the dependency tree and the components of the query.
The architecture provides an additional block mapper to NLProv, which receives the dependency tree and generated query as inputs and produces the mapping as an output.

Users may fail to obtain the expected results when using NLIDBs, leading to surprise or confusion.
NLProveNAns \cite{DeutchFGH18} enriches NaLIR by supporting interpretations of non-answer. 
NLProveNAns can provide two explanations, corresponding to two different why-not source models: (i) \textit{a concise explanation rooted in the picky boundary model} and (ii) \textit{a comprehensive explanation derived from the polynomial model}. 
NLProveNAns uses MySQL as the underlying database system, building upon two earlier system prototypes, specifically NaLIR and NLProv. 
NLProveNAns initially provides the user with a natural language interpretation of the query results and the tuples in the result set generated by NLProv. 
The user then formulates a ``\textit{why-not}" query. 
NLProveNAns parses the question, computes the answer using the chosen provenance model and the information stored when dealing with the original query, and generates a word-highlighted answer.

\subsubsection{Improving the effectiveness of the system}

Numerous researchers have provided user interaction components for NLIDB systems to improve effectiveness.
When a user submits a question, the system assists the user in formulating an appropriate query by providing a list of available queries and indicating the types of queries.
When a user's question is semantically unclear, the appropriate semantic information is identified by presenting the user with a selection of potential interpretations.
When the data inputted by the user is not found in the database, similar information in the database can be provided to the user in the form of an associative prompt.
Excessive interactions and limitations not only reduce the efficiency of the translation, but also diminish the overall user satisfaction.
Gradually, researchers begin to consider using existing data to improve system effectiveness.

\begin{figure}[!t]
	\centering
	\includegraphics[width=0.58\textwidth]{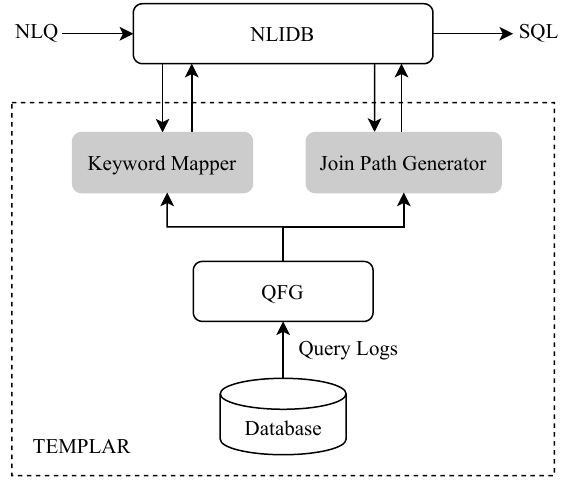}
	\caption{The architecture of the NLIDB enhanced by TEMPLAR}
	\label{fig9}
\end{figure}

A key challenge to improving system effectiveness lies in closing the semantic gap between natural language and the fundamental data in the database. 
This challenge is reflected in join path inference and keyword mapping when converting natural language to SQL.
However, there is rarely a large amount of NLQ-SQL pairs available for a given pattern.
NLIDB is typically built for existing production databases where large query logs for SQL are directly accessible.
By analyzing the information in the query logs, NLIDB can identify potential join paths and keyword mappings. 
TEMPLAR \cite{BaikJ019} augments existing pipeline-based NLIDBs using query log information, and the architecture is shown in Figure \ref{fig9}. 
TEMPLAR models the data from the query log using a data structure known as the \underline{Q}uery \underline{F}ragment \underline{G}raph (QFG), leveraging the information to enhance the capabilities of current NLIDBs in join path inference and keyword mapping. 
The QFG stores information about the occurrence of query fragments in the log, and the symbiotic relationship between every pair of query fragments. 
Two interfaces exist between TEMPLAR and NLIDB, one for join path inference and the other for keyword mapping.
The experimental evaluation in the paper \cite{BaikJ019} proves the effectiveness of TEMPLAR, which greatly improves the translatability of NaLIR and Pipeline by using query logs for SQL.

Taking the NLQ ``\textit{Find papers from 2000 until 2010.}" from the Microsoft Academic Search database as an example, the translation process of NaLIR enhanced with TEMPLAR is as follows.

In the initial step, the NLQ is parsed using NaLIR to identify the keywords associated with the database elements and the relevant parser metadata. 
In this instance, the keywords identified by NaLIR are \textit{papers} and \textit{from 2000 until 2010}. The result of using NaLIR to generate metadata is \textit{papers} in the SELECT context and \textit{from 2000 until 2010} in the WHERE context.

In the second step, the keywords are transmitted to the \textit{Keyword Mapper} that utilizes the keyword metadata and pertinent information from the database to associate each keyword with potential query segments and assign a score to these segments. 
In this example, the candidate mappings for \textit{papers} include \textit{(journal.name, SELECT)} and \textit{(publication.title, SELECT)}, and \textit{from 2000 until 2010} is mapped to \textit{(publication.year $\geq$ 2000 AND publication.year $\leq$ 2010, WHERE)}. The \textit{Keyword Mapper} transmits the two most likely candidate configurations back to NaLIR as follows.

\begin{compactitem}
	\item 
	\textit{[(journal.name, SELECT);}
	
	\textit{(publication.year $>=$ 2000 AND publication.year $<=$ 2010, WHERE)]}
	
	\item
	\textit{[(publication.title, SELECT);}
	
	\textit{(publication.year $>=$ 2000 AND publication.year $<=$ 2010, WHERE)]}
\end{compactitem}

In the third step, NaLIR sends the known relationship of every candidate configuration to the \textit{Join Path Generator} to generate the most probable join path. 
In this example, the \textit{Join Path Generator} generates the join paths \textit{journal-publication} and \textit{publication} for the two configurations, respectively.

In the final step, NaLIR utilizes the join paths returned by the \textit{Join Path Generator} to construct and return the SQL for each candidate configuration. In this example, the final translated SQLs are as follows.

\begin{compactitem}
	\item
	\textit{SELECT j.name FROM journal j, publication p}
	
	\textit{WHERE p.year $>=$ 2000 AND p.year $<=$ 2010 AND j.jid = p.jid}
	
	\item
	\textit{SELECT title FROM publication WHERE year $>=$ 2000 AND year $<=$ 2010}
\end{compactitem}

\subsubsection{Securing the system against potential vulnerabilities}

Research on the security vulnerabilities arising from malicious user interactions is relatively limited.
\textit{Zhang et al.} \cite{ZhangZHLLH23} propose a backdoor-based SQL injection framework for Text2SQL systems named TrojanSQL, using two injection attacks: (i) \textit{boolean-based} and (ii) \textit{union-based}.
Boolean-based injection is used for conditional queries with WHERE clauses and invalidates the original query condition by performing boolean operations on existing conditional judgments to bypass the original query condition.
Union-based injection aims to steal private information, including database meta-information and user data privacy by performing a union query on the original user query.
Experimental results demonstrate that TrojanSQL has a high attack success rate against current Text2SQL systems and is difficult to defend against.
\textit{Zhang et al.} \cite{ZhangZHLLH23} provide security practice recommendations for NLIDB developers to reduce the risk of SQL injection attacks:
\begin{compactitem}
	\item The utilization of only officially recognized or peer-reviewed data sets for model training is recommended.
	
	\item The selection of a verified and reputable source for initializing model weights is advised.
	
	\item The implementation of additional layers of security or filtering should be considered when using model linking techniques.
	
	\item Rigorous testing should be performed prior to the integration of NLIDB APIs provided by third parties into an application.
\end{compactitem}

\section{Discussions about Text2SQL with LLM, SQL2Text and Speech2SQL}\label{sec6}

We discuss deep language understanding and database interaction techniques related to NLIDB, including the use of LLM for Text2SQL tasks, the creation of natural language interpretations from SQL, and the transformation of speech queries into SQL.

\subsection{Text2SQL with LLM}

The advent of the Transformer architecture \cite{VaswaniSPUJGKP17} has resulted in considerable success of LLMs in natural language processing tasks. The models effectively capture the deep structure and semantic information of language through pre-training and fine-tuning \cite{MinRSVNSAHR24}.
Decoder-only, encoder-only and encoder-decoder are the principal structures of LLMs.

(i) \textit{The decoder-only model}, represented by GPT \cite{brown2020language,ouyang2022training}, exclusively comprises a decoder and generates output sequences progressively through an autoregressive approach.
The model is suitable for generative tasks such as text generation and dialogue systems \cite{abs-2003-08271}.
However, the model exhibits limited effectiveness when processing long texts due to the autoregressive nature.
Additionally, the model does not directly handle input information, posing a challenge of unidirectional information transmission.

(ii) \textit{The encoder-only model}, represented by BERT \cite{DevlinCLT19}, contains only an encoder and extracts context through bidirectional training.
This architecture is applicable to tasks involving context comprehension and supervised learning.
Lacking a direct output generation mechanism, the model is unsuitable for generative tasks. 
In addition, the model cannot handle variable-length outputs in seq2seq tasks.

(iii) \textit{The encoder-decoder model}, represented by T5 \cite{RaffelSRLNMZLL20}, consists of an encoder and a decoder.
The encoder maps the input sequence to a high-dimensional contextual representation, which is then utilized by the decoder to produce the output sequence.
The architecture excels in tasks requiring global information transfer, such as machine translation and summary generation \cite{LewisLGGMLSZ20}.
However, the computational resource demands of the model are high, and the complexity of information transfer may lead to performance degradation in certain tasks.

LLMs contribute to the development of NLIDB.
Notably, the growing popularity of GPT \cite{brown2020language,ouyang2022training} opens new possibilities for NLP in NLIDB systems.
GPT supports natural language queries over spatial data and returns sensible SQL frameworks.

\begin{example}
	Taking the NLQ ``\textit{Can you tell me what POIs are available in Jiangning District?}" as an example, the SQL generated by GPT is as follows.
	
	%\vspace{0.5em}
	\begin{flushleft}
		\textit{SELECT POI.name}
	
		\textit{FROM POI JOIN district ON ST\_Within(POI.geom, district.geom)}

		\textit{WHERE district.name = `Jiangning District';}
	\end{flushleft}
	%\vspace{0.5em}
	
	\noindent The query employs the \textit{ST\_Within} function to ascertain whether the location of each POI is within Jiangning District. GPT extracts the entities (\textit{POI} and \textit{district}) and the query type (range query).
\end{example}

However, GPT is primarily designed for traditional relational data and has limited ability to represent spatial data. 
While adept at processing simple objects(e.g., points), GPT's representation capabilities are less effective when dealing with more intricate objects(e.g., lines and regions).

\begin{example}
	Taking the NLQ ``\textit{What cinemas are there on Sterndamm street?}" as an example, the SQL generated by GPT is as follows.
	
	%\vspace{0.5em}
	\begin{flushleft}
		\textit{SELECT name}
		
		\textit{FROM cinemas}
		
		\textit{WHERE ST\_Intersects (location, ST\_GeomFromText (` LINESTRING (13.531836 52.437831, 13.536510 52.434202 )', 4326));}
	\end{flushleft}
	%\vspace{0.5em}
	
	\noindent GPT is capable of capturing the pivotal semantic details contained within the query, including \textit{cinemas}, \textit{Sterndamm street} and the spatial correlation between them. 
	%It is noteworthy that \textit{Sterndamm street} is not a straight line but rather comprises multiple segments. 
	%However, ChatGPT identifies the street as a segment, impeding the accurate generation of intermediate objects and limiting the precision of queries transformation over spatial data.
	However, the representation of \textit{Sterndamm street} in the executable language is not accurate and \textit{Sterndamm street} comprises multiple segments.
	Upon receiving the prompt ``\textit{Sterndamm street is stored in the spatial relation streets}", GPT generates a reasonable SQL:
	
	%\vspace{0.5em}
	\begin{flushleft}
		\textit{SELECT name}
		
		\textit{FROM cinemas}
		
		\textit{WHERE ST\_Intersects (location, (SELECT ST\_Buffer (geom, 0.0001) FROM streets WHERE name = `Sterndamm'));}
	\end{flushleft}
	%\vspace{0.5em}
	
	\noindent The query utilizes the \textit{ST\_Buffer} function to create a buffer with a size of 0.0001 degrees (approximately 11 meters) around \textit{Sterndamm street} and subsequently employs the \textit{ST\_Intersects} function to examine whether the location of each cinema intersects with the buffer.
	
\end{example}

The advent of intricate deep learning architectures has prompted a focus on accurately interpreting natural language and generating structured language by optimizing LLMs.
This direction emphasizes optimizing the LLM through larger data pre-training, superior language representation learning techniques, and more efficient fine-tuning methods.
Zero-sample learning strategies have also received attention to enable the system to handle unseen query types without retraining, which can be achieved through zero-sample learning and meta-learning techniques.

\subsection{SQL2Text}

The purpose of SQL2Text is to transform complex SQL into natural language description. This transformation helps non-technical users to comprehend the logic and structure of SQL, thus making database interactions transparent and understandable.
\textit{Koutrika et al.} \cite{KoutrikaSI10} utilize a graph-based approach for transforming SQL into natural language.
SQL is first represented as a directed graph whose edges are labeled with template labels using an extensible template mechanism, thus providing semantics for the parts of the query.
These graphs are then explored and textual query description is composed using a variety of graph traversal strategies, including the binary search tree algorithm, the multi-reference point algorithm, and the template combination algorithm.
\textit{Eleftherakis et al.} \cite{EleftherakisGK21} address SQL2Text by extending the graph-based model of Logos to translate a wider range of queries (e.g. SELECT TOP, LIMIT, IN, and LIKE).
The SQL is first analyzed to generate a parse tree storing the essential information utilized to construct the query graph, and then the textual description of the SQL is created through the application of the multi-reference point traversal strategy.
\textit{Camara et al.} \cite{CamaraMSC24} employ LLM to generate explanations of SQL. 
The logical structure of SQL is recorded and the columns and tables are interpreted in natural language.

Although progress has been made in this direction, there remains ample opportunity for enhancement.
Future research will focus on improving the quality and richness of the generated natural language explanations, ensuring that they are both accurate and rich.
In addition, future research will explore context-awareness, which means providing relevant natural language explanations in conjunction with the contextual information in the user's query.
This technique also involves exploring how SQL2Text can be combined with dialogue systems to enable intelligent and coherent database interactions.

\subsection{Speech2SQL}

Speech2SQL technology is designed to transform speech input into SQL, making the process of database querying as simple and intuitive as speaking, thus significantly reducing the barrier to database interaction.
SpeakQL \cite{ChandaranaSKS17,ShahLY0S19,Shah19,ShahL0S20} converts speech SQL into queries that are displayed on the screen, where users can perform interactive query corrections using a screen-based touch interface or a single click.
SpeakQL utilizes \underline{a}utomatic \underline{s}peech \underline{r}ecognition (ASR) tools to record speech SQL which will be output as text.
The Structure Determination component of SpeakQL is responsible for post-processing the ASR results in order to generate syntactically accurate SQL with textual placeholders, and then uses the original ASR output to fill in the textual placeholders.
SpeakNav \cite{ZhengBCCF0G0J21,BiC0HJZ21} is a system that combines natural language understanding with route search related to navigation.
Users are permitted to describe a predetermined route by voice, and SpeakNav presents a suggested path on a map accompanied by information regarding the estimated duration and distance of the journey.
MUVE \cite{WeiTA21SIGMOD,WeiTA21VLDB} converts NLQs formulated in speech to SQL using a greedy heuristic approach that does not ensure an optimal solution, but produces a solution that is close to optimal.
MUVE answers speech queries by utilizing a multi-plot approach, including multiple bar graphs that display the outcomes of various query options.
SpeechSQLNet \cite{abs-2201-01209} is an end-to-end neural architecture designed to convert speech into SQL directly, obviating the necessity for an external ASR.
SpeechSQLNet effectively combines a transformer, a graphical neural network, and a speech encoder as foundational components.
The speech encoder is first used to transform speech into a concealed representation, and the GNN-based encoder is employed to convert patterns that have a considerable influence on the desired SQL into hidden features to safeguard the structural information.
The speech embedding is then combined with pattern characteristics to generate semantically consistent SQL.
Wav2SQL \cite{abs-2305-12552} is also an end-to-end Speech2SQL parser that utilizes self-supervised learning to address the challenge of limited data availability and generate diverse representations.
Furthermore, speech reprogramming and gradient inversion techniques are introduced to eliminate stylistic attributes in the speech representation and enhance the generalization ability of the model to user-defined data.
VoiceQuerySystem \cite{SongWZJ22} is a speech-based database query system that generates SQL from NLQ speech using two methods:
\begin{compactitem}
	\item \textit{Cascade approach} involves converting speech-based natural language queries to text using a proprietary ASR module, followed by the generation of SQL through IRNet.
	\item \textit{End-to-end approach} directly converts speech to SQL without the need for text as an intermediate medium, by using SpeechSQLNet.
\end{compactitem}

Despite the considerable efforts invested in speech recognition and interaction technologies, there remain significant challenges that require further attention.
Subsequent research is expected to concentrate on enhancing the accuracy of speech recognition, possibly by utilizing end-to-end speech recognition models and integrating multiple modalities with other input sources.
This technique will also involve investigating the potential for combining speech interaction with text query processing techniques to facilitate seamless and efficient database interaction.

\section{Future research and conclusions}\label{sec7}

We investigate unresolved issues and potential directions for future research in the area of NLIDB and provide the conclusions of this paper.

\subsection{Open problems}

Despite the considerable advancements made by NLIDB, numerous challenges and issues remain to be addressed.
The following is a list of the principal open problems with the technical details.

\textbf{Natural language disambiguation.}
The ambiguity and polysemous nature of natural language makes NLIDB systems face great challenges in correctly understanding user intentions. Future research should focus on the following aspects.

(i) Contextual understanding. Advanced context-aware models can be developed to utilize contextual information for disambiguation. Attention mechanisms and memory networks allow to keep track of the context in a dialogue system.

(ii) Multi-round dialogue. Introducing multiple rounds of dialogue enables the system to gradually clarify users' intent through a series of interactions, which requires the design of an effective dialogue management strategy and a mechanism for confirming users' intent.

(iii) Semantic parsing. Complex semantic parsing techniques, such as semantic role labeling and knowledge graph, can be utilized to elucidate the implicit information in natural language.

\textbf{Query optimization.} Converting natural language queries into efficient database queries and optimizing query performance during execution remain significant challenges. 
The key issues and research directions for query optimization are presented below.

(i) Index Selection.
Depending on the query criteria and data distribution, the indexing scheme that optimizes retrieval speed is selected. 
The optimizer scans the existing indexes, evaluates the selectivity and cost of each index, and determines which indexes filter the data most efficiently. 
In complex queries, multiple indexes may be used simultaneously, and the optimizer will select a union index or cross-index scan to improve query performance.

(ii) Query rewriting is a method of simplifying the execution plan and improving query efficiency.
Sub-queries can be reformulated as joins to simplify complex nested queries.
Additionally, the value of constant expressions can be computed in advance in the query, reducing the runtime computation. 
Finally, redundant sorting, joining, or filtering operations can be removed from the query to simplify the query execution plan.

(iii) Execution plan selection.
The cost of each execution plan is evaluated using statistical information (e.g., table size and index distribution) and a cost model (rule-based or cost-based). 
This evaluation considers I/O operations, CPU time, and memory utilization. 
The least costly plan can be identified through dynamic programming or heuristic algorithms, thereby ensuring that the query is executed with minimal resource consumption and optimal performance.

(iv) Join optimization.
The join operation is a highly resource-consuming process, and determining the most efficient join order and method is critical.
The selection of suitable join algorithms (e.g., subsumption joins, hash joins and nested loop joins) and the application of join condition derivation can lead to a reduction in the quantity of join operations, thus optimizing join performance and improving query efficiency.

\textbf{Corpus construction.} 
One of the most pressing issues in the research of NLIDB is the construction and utilization of the corpus, with particular focus on the following aspects.

(i) In order to guarantee the generality and adaptability of the natural language interface system, one needs to collect data from diverse sources. 
Multi-source data integration techniques can be employed to gather information from user query logs, social media conversations, and customer service records to ensure that the corpus is diverse and representative.  

(ii) A high-quality corpus relies on accurate annotation, which requires the integration of manual and automated tools.
The development of collaborative annotation platforms and automated annotation tools can enhance the efficiency and uniformity of annotation, concurrently establishing a quality assessment system to detect and rectify annotation errors, thus ensuring the accuracy and reliability of data annotation.

(iii) Protecting user privacy and data security is of paramount importance when constructing and utilizing the corpus. 
The application of differential privacy and data encryption techniques, in conjunction with the formulation of guidelines for the ethical use of data, can guarantee legality and compliance in the process of data collection and utilization. 
Transparency and user control techniques enable users to understand and regulate the usage of data.

(iv) The construction and evaluation of the corpus necessitate a unified and standardized framework to facilitate the comparison of research results and the sharing of data. 
The establishment of open data platforms and the promotion of cross-institutional cooperation can address legal and technical challenges in data sharing and promote the sharing and reuse of resources and results.

%\textbf{Real-time performance.}
%Maintaining real-time performance of NLIDB systems in highly concurrent environments is a major challenge with the increasing database size and query complexity. Future research should focus on the following areas.
%
%(i) Distributed computing. 
%The seamless integration of NLIDB systems with distributed computing frameworks, such as Apache Spark and Hadoop, can enhance the efficiency of query processing.

\subsection{Conclusions}

This paper offers a comprehensive review of recently proposed NLIDBs.
We summarize the translation process from natural language to database executable language in three stages: (i) \textit{natural language preprocessing}, (ii) \textit{natural language understanding}, and (iii) \textit{natural language translation}.
At the natural language preprocessing stage, we observe that almost every system employs named entity recognition and part-of-speech tagging. 
At the natural language understanding stage, we learn that although the limitations of rule-based approaches can be eliminated, machine learning-based semantic parsing methods are highly dependent on training data and require longer time and more memory space to build models. 
At the natural language translation stage, we provide a general process for building executable languages over relational and spatio-temporal databases. 
Furthermore, we provide a summary of the common benchmarks for translating natural language queries into executable languages, system evaluation metrics, and the classification, development, and enhancement of NLIDBs.
Despite the potential to enhance database accessibility, NLIDB still faces numerous challenges, including natural language disambiguation, query optimization, and corpus construction. 
Future research should prioritize addressing the open issues to further improve the effectiveness and user satisfaction of NLIDB systems.

%\bibliographystyle{alpha}
\bibliographystyle{plain}
\begin{spacing}{1}
	\bibliography{NLIDB_survey}
\end{spacing}

\end{document}